\documentclass[fleqn,usenatbib]{mnras}

\usepackage[T1]{fontenc}

\DeclareRobustCommand{\VAN}[3]{#2}
\let\VANthebibliography\thebibliography
\def\thebibliography{\DeclareRobustCommand{\VAN}[3]{##3}\VANthebibliography}

\usepackage{graphicx}	
\usepackage{amsmath}	
\usepackage{amssymb}	
\usepackage{multirow}   
\usepackage{comment}
\makeatletter
\newcommand\thefontsize[1]{{#1 The current font size is: \f@size pt\par}}
\makeatother

\DeclareTextFontCommand{\textttup}{\normalfont\ttfamily}
\makeatletter
\newcommand{\printexternalcurrentfont}{%
  \expandafter\format@externalcurrentfont\fontname\font:\@nil
}
\def\format@externalcurrentfont#1:#2\@nil{%
  \textttup{\@ifnextchar"{\@gobble}{}#1}%
}
\makeatother

\usepackage{newtxtext,newtxmath}

\defcitealias{Scott+17}{S17}
\defcitealias{McDermid+15}{M15}
\defcitealias{Bernardi+19}{B19}
\defcitealias{Peng+10}{P10}

\title[{$[\alpha/\text{Fe}]$-$\sigma$ Relations in the SGS}]{The SAMI Galaxy Survey: Trends in $[\alpha/\text{Fe}]$ as a Function of Morphology and Environment}

\author[P. J. Watson et al.]{Peter J. Watson,$^{1}$\thanks{E-mail: peter.watson2@physics.ox.ac.uk}
Roger L. Davies,$^{1}$
Sarah Brough,$^{2,3}$
Scott M. Croom,$^{2,4}$
Francesco D'Eugenio,$^{5}$
\newauthor Karl Glazebrook,$^{2,6}$
Brent Groves,$^{7}$
\'Angel R. L\'opez-S\'anchez,$^{2,8,9}$
Jesse van de Sande,$^{2,4}$
\newauthor Nicholas Scott,$^{2,4}$
Sam P. Vaughan,$^{4}$
Jakob Walcher,$^{10}$
Joss Bland-Hawthorn,$^{4}$
Julia J. Bryant,$^{2,4,11}$
\newauthor Michael Goodwin,$^{8}$
Jon S. Lawrence,$^{12}$
Nuria P. F. Lorente,$^{8}$
Matt S. Owers,$^{2,9,13}$
and Samuel Richards$^{14}$
\\
$^{1}$ Sub-Department of Astrophysics, Department of Physics, University of Oxford, Denys Wilkinson Building, Keble Rd., Oxford OX1 3RH, UK\\
$^{2}$ ARC Centre of Excellence for All Sky Astrophysics in 3 Dimensions (ASTRO 3D)\\
$^{3}$ School of Physics, University of New South Wales, NSW 2052, Australia\\
$^{4}$ Sydney Institute for Astronomy (SIfA), School of Physics, University of Sydney, NSW 2006, Australia\\
$^{5}$ Sterrenkundig Observatorium, Universiteit Gent, Krijgslaan 281 S9, B-9000 Gent, Belgium\\
$^{6}$ Centre for Astrophysics \& Supercomputing, Swinburne University of Technology, Victoria 3122, Australia\\
$^{7}$ Research School of Astronomy \& Astrophysics, Australian National University, Mt Stromlo Observatory, Cotter Rd, Weston Creek, ACT 2611, Australia\\
$^{8}$ Australian Astronomical Optics, Macquarie University, 105 Delhi Rd, North Ryde, NSW 2113, Australia\\
$^{9}$ Department of Physics and Astronomy, Macquarie University, NSW 2109, Australia\\
$^{10}$ Leibniz-Institut f\"ur Astrophysik Potsdam (AIP), An der Sternwarte 16, D-14482 Potsdam, Germany\\
$^{11}$ Australian Astronomical Optics, AAO-USydney, School of Physics, Building A28, University of Sydney, NSW 2006, Australia\\
$^{12}$ Australian Astronomical Optics - Macquarie, Macquarie University, NSW 2109, Australia\\
$^{13}$ Astronomy, Astrophysics and Astrophotonics Research Centre, Macquarie University, Sydney, NSW 2109, Australia\\
$^{14}$ SOFIA Science Center, USRA, NASA Ames Research Center, Building N232, M/S 232-12, P.O. Box 1, Moffett Field, CA 94035-0001, USA\\
}

\date{Accepted XXX. Received YYY; in original form ZZZ}

\pubyear{2020}

\begin{document}
\label{firstpage}
\pagerange{\pageref{firstpage}--\pageref{lastpage}}
\maketitle

\begin{abstract}
We present a new set of index-based measurements of $[\alpha/\text{Fe}]$ for a sample of 2093 galaxies in the SAMI Galaxy Survey.
Following earlier work, we fit a global relation between $[\alpha/\text{Fe}]$ and the galaxy velocity dispersion $\sigma$ for red sequence galaxies, $[\alpha/\text{Fe}]=(0.378\pm0.009)\rm{log}_{10}\left(\sigma/100\right)+(0.155\pm0.003)$.
We observe a correlation between the residuals and the local environmental surface density, whereas no such relation exists for blue cloud galaxies.
In the full sample, we find that elliptical galaxies in high-density environments are $\alpha$-enhanced by up to $0.057\pm0.014$\,dex at velocity dispersions $\sigma<100$\;km\,s$^{-1}$, compared with those in low-density environments.
This $\alpha$-enhancement is morphology-dependent, with the offset decreasing along the Hubble sequence towards spirals, which have an offset of $0.019\pm0.014$\,dex.
At low velocity dispersion and controlling for morphology, we estimate that star formation in high-density environments is truncated $\sim1$\,Gyr earlier than in low-density environments.
For elliptical galaxies only, we find support for a parabolic relationship between $[\alpha/\text{Fe}]$ and $\sigma$, with an environmental $\alpha$-enhancement of at least $0.03$\,dex.
This suggests strong contributions from both environment and mass-based quenching mechanisms.
However, there is no evidence for this behaviour in later morphological types.
\end{abstract}




\section{Introduction}

\subsection{Background} \label{sec:background}

Analysing the star-formation history for any given galaxy is not an insignificant task. 
For the overwhelming majority of galaxy observations so far, it is not possible to resolve individual stars, limiting any analysis to the integrated light over the observable area. 
The simplest method of differentiating stellar populations is using galaxy colours.
By observing resolved stars in specific filters, such as those in the ubiquitous UBV photometric system of \cite{Johnson+53}, galaxies could be classified according to their colour.
This provided a basis for the core component of galaxy evolution analysis: stellar population synthesis.
By combining various properties of resolved stars with stellar evolutionary tracks and an initial mass distribution, it was possible to synthesise the UBV colours of a model galaxy.
Spectrophotometric models such as those of \cite{Searle+73} provided a straightforward reference for the age, star-formation rate, and mass distribution of extragalactic stellar populations.
However, the triple degeneracy of dust reddening, stellar age, and metallicity severely limits the use of colour for population analysis.

Far stronger constraints can be placed on the content of galaxies through the use of spectroscopy. 
Within the restframe optical and near-infrared (NIR) region, there are a multitude of strong spectral features.
Emission lines can be used to derive star-formation rates, and gas kinematics, whilst the absorption lines reveal a wealth of information on the stellar content.
The absorption lines in the wavelength region 4000-6000\,\AA\ are largely insensitive to dust attenuation \citep{MacArthur+05}, and can be measured easily and calibrated to a common system, with one of the most widely used being the the Lick/IDS system, thoroughly detailed in \cite{Worthey+97}.

The Lick Index System is a set of 25 absorption-line features, including 5 Balmer absorption-line features, at optical wavelengths defined by the Lick group over a series of papers \citep{Burstein+84,Faber+85,Gorgas+93,Worthey+94,Worthey+97,Trager+98}.
By combining features that are separately sensitive to age and metallicity, the degeneracy between these two variables can be significantly reduced, a marked improvement over the use of UBV colours.
Although non-trivial, the variation of some of these absorption-line indices with individual elemental abundances can also be calculated.
\cite{Thomas+03} were among the first to produce model predictions of Lick indices with variable elemental abundance, based on synthesised stellar populations.
By comparison with index measurements from galaxy spectra, these models and others like them, such as \cite{Schiavon+07} and \cite{TMJ+11}, allow a straightforward determination of the luminosity-weighted constituent single stellar population (SSP) parameters.
In most cases, these models predict index measurements as a function of age, metallicity, and the $\alpha$-element abundance.

Alongside the Lick Index System, there are many alternative methods for extracting the integrated star-formation history from spectra.
One attractive concept is that of full-spectrum fitting, where models of the spectral energy distribution (SED) are compared against the observed galaxy spectrum, allowing a more direct determination of properties such as the initial stellar mass function (IMF).
Various implementations exist, such as \textsc{bagpipes} from \cite{Carnall+18}, based on the models of \cite{Bruzual+03}, or \textsc{PyStaff} from \cite{Vaughan+18b}, which utilises the models of \cite{Conroy+18}, and techniques first implemented in \textsc{ppxf} \citep{Cappellari+17}, which itself is often used for stellar population analysis.

The significant advantage of full-spectrum fitting over using only indices is the ability to fit for individual elemental abundances simultaneously, without limiting the analysis to pre-selected regions of interest, although this comes with a correspondingly steep computational cost, and very stringent requirements for the wavelength and flux calibration.
The variation in techniques used to create the underlying SED models can also strongly influence the results, with models such as \cite{Conroy+18} ``correcting'' an empirical spectra to some arbitrary elemental abundance via individual elemental response functions, whereas the self-consistent models of \cite{Walcher+09} and \cite{Vazdekis+15} include the additional constraint of matching the elemental abundances of the isochrones.
Thus, Lick indices still represent arguably the best compromise between the wealth of data that can be extracted from an individual spectra, and the computational cost of doing so.
This makes them ideal for larger galaxy surveys, where the cost of full spectrum fitting becomes prohibitively expensive.

\subsection{Previous work} \label{sec:previous_work}

The most common abundance pattern to be measured is the $\alpha$-element abundance, $[\alpha/\text{Fe}]$.
These are elements whose most abundant isotope comprises an integer number of $\alpha$ particles, such as C, O, Mg, Ca, and Ti. 
They are predominantly formed in massive stars ($M\geq8M_{\odot}$) during the oxygen and silicon burning processes, the two penultimate stages before the stellar core collapses, giving rise to a Type II supernova.
This contrasts with Type Ia supernovae, which mainly form Fe-peak elements, such as Mn, and Co.
Thus, $[\alpha/\text{Fe}]$ for a given star measures the relative contributions of Type Ia and Type II supernovae to the ISM, at the point in time when that star formed.
If star formation continues over long timescales, the relative contribution from SNe Ia becomes more significant, and so $[\alpha/\text{Fe}]$ for any newly formed stars trends towards solar values \citep{Greggio+83}.
Over a large stellar population, many factors can affect the integrated $[\alpha/\text{Fe}]$, such as the delay-time-distribution of SNIa, duration of star formation, and variations in the IMF.

Previous spectroscopic studies into $[\alpha/\text{Fe}]$ in early-type galaxies (ETGs), such as \cite{Trager+00}, have revealed that there is a strong empirical correlation with both stellar velocity dispersion $\sigma$, and stellar mass.
This clear relation makes it an ideal test for cosmological simulations and semi-analytic models of galaxy formation.
Using the EAGLE project, \cite{Segers+16} have shown that the correlation can be reproduced for simulated galaxies with stellar masses of $M_*>10^{10.5}M_\odot$.
They attribute this to quenching from active galactic nuclei (AGN) feedback, which suppresses star formation in more massive galaxies at earlier times, and show that the observations are inconsistent with simulations in the absence of this feedback.
In addition, \cite{Calura+18} also posit that ``fly-by'' harrassments may be a significant factor in causing brief bursts of star formation.
With current work finding a consistent SNIa delay-time-distribution across models and observations via multiple techniques \citep{Maoz+12, Walcher+16}, and the assumption that there exists some universal IMF, the general consensus is thus that $[\alpha/\text{Fe}]$ more directly correlates to the star-formation timescale, as demonstrated in \cite{de_la_Rosa+11}.

However, there is still considerable uncertainty over the behaviour of low stellar mass galaxies, and in different environments.
Having a clear understanding of star-formation timescales as a function of these variables can highlight the relative importance of different quenching mechanisms, \textit{e.g.} ram-pressure stripping \citep{Gunn+72}, strangulation \citep{Larson+80}, or tidal forces \citep{Dekel+03}.
Looking at morphologically selected ETGs, \cite{Thomas+10} found scaling relations with $[\alpha/\text{Fe}]$ were not sensitive to the environmental densities, although $\sim$10\% of galaxies displayed signs of ongoing star formation, and were correspondingly less $\alpha$-enhanced by approximately 0.1\,dex.
\cite{Annibali+11} instead found that dwarf galaxies, selected by $1.6<\rm{log}_{10}(\sigma)<2$, in high-density environments (HDEs) were enhanced in $[\alpha/\text{Fe}]$ by 0.22\,dex compared to galaxies in low-density environments (LDEs).
\cite{La_Barbera+14} looked at the variation of SSP parameters in both central and satellite ETGs in high and low density environments.
They found no environmental dependence for $[\alpha/\text{Fe}]$ in satellite galaxies, but a consistent $\alpha$-enhancement of $\sim$0.025\,dex for central galaxies in LDEs over their counterparts in HDEs.

\citet[][hereafter:\ \citetalias{Scott+17}]{Scott+17}, looking at a volume-limited sample separated by local environmental density, found an increase of $\Delta[\alpha/\text{Fe}]=0.14\pm0.02$ from LDEs to HDEs, in galaxies with stellar masses $M_*>10^{10.5}M_\odot$.
\citetalias{Scott+17} also investigated correlations with morphology, and found ETGs were $\alpha$-enhanced by $0.07\pm0.03$ dex over spirals with the same stellar mass.
They also found that this dependence was primarily driven by galaxy size, which was anti-correlated with environmental density.
\citet[][hereafter:\ \citetalias{McDermid+15}]{McDermid+15} fitted their own sample, consisting of ETGs, to a common $[\alpha/\text{Fe}]$-$\sigma$ relationship. 
Separating these galaxies into Virgo cluster members and non-members, they were able to show that the residuals for each sample were unlikely to be drawn from the same parent distribution. 

The primary limitation on many of these studies has been the small sample size, particularly when comparing galaxies at low velocity dispersion.
The largest surveys in this area are also typically single-fibre based, which can introduce unwanted aperture effects such as those explored in \citetalias{McDermid+15}.
Through the use of integral field spectroscopy (IFS), we can scale the effective aperture to match each galaxy, reducing aperture bias.
There exist several large IFS surveys to date, including ATLAS$^{\text{3D}}$ \citep{Cappellari+11}, the CALIFA Survey \citep{Sanchez+12}, and SDSS-IV MaNGA \citep{Bundy+15}.
Here, we present stellar population measurements, using aperture spectra drawn from the SAMI (Sydney-AAO Multi-object Integral field spectrograph) Galaxy Survey, hereafter referred to as the SGS. 
In Section \ref{sec:Data} we describe the SGS in more detail, including the sources of the ancillary data used.
The method used for extracting the luminosity-weighted SSP-equivalent parameters is detailed in Section \ref{sec:method}.
We present our results, the dependence of $[\alpha/\text{Fe}]$ on both environment and morphology, in Section \ref{sec:environmental_dependence}.
Finally, we discuss the implications of our research and conclude in Sections \ref{sec:discussion} and \ref{sec:conclusions}.
Throughout this paper we assume a $\Lambda$CDM cosmology, with $\Omega_{\rm{m}}=0.3$, $\Omega_\Lambda=0.7$, and $H_0=70$\,km\,s$^{-1}$Mpc$^{-1}$.

\section{Data}
\label{sec:Data}

\subsection{SAMI Galaxy Survey} 
\label{sec:SAMI_galaxy_survey}

The SAMI instrument and survey design are detailed extensively in both \cite{Croom+12} and \cite{Bryant+15}.
The instrument comprises 13 IFUs (the \textit{hexabundles}), which can be deployed over a 1-degree diameter field of view, each with an individual field of view of 15 arcsec \citep{Bland_Hawthorn+11, Bryant+14}.
The IFUs are mounted at the prime focus of the Anglo-Australian Telescope (AAT), and each consists of 61 individual fibres.
Observations are dithered to create data cubes with a 0.5-arcsec spaxel size.
All 819 fibres (including 26 allocated to blank sky observations for calibration purposes) are fed into the AAOmega spectrograph \citep{Saunders+04, Smith+04, Sharp+06}.
This is composed of a blue arm, with spectral resolution $R\sim1800$ over the wavelength range 3750-5750\,\AA, and a higher resolution red arm, with wavelength coverage 6300-7400\,\AA\ and $R\sim4300$ \citep{van_de_Sande+17a}.

The SGS consists of 3426 observations of 3068 unique galaxies, available as part of public Data Release 3 \citep{Croom+21}.
The survey spans a redshift range $0.004<z<0.115$, and a stellar mass range $M_*\sim10^7$ to $10^{12}\,M_{\odot}$.
Field and group galaxies were drawn from the Galaxy And Mass Assembly (GAMA) survey \citep{Driver+11}, with the selection being volume-limited in each of four stellar mass cuts.
An additional sample of cluster galaxies was drawn from the survey of eight low-redshift clusters in \cite{Owers+17}, to extend the environmental sampling.

\subsection{Ancillary data}
\label{subsec:ancillary_data}

Throughout our analysis, we make extensive use of additional measurements by other members of the SAMI team. 
These include circularised effective radii ($r_e$), measured using the Multi Gaussian Expansion  \citep[MGE,][]{Emsellem+94} algorithm of \cite{Cappellari+02} and photometric fits by \cite{D'Eugenio+21}.
Optical morphological classifications are taken from SAMI Public Data Release 3 \citep{Croom+21}, following the method of \cite{Cortese+16}, where galaxies were designated as one of four types (Ellipticals, S0s, early- and late-type spirals).
This was based on visual inspection of colour images by $\sim$10 SAMI team members, taken from either SDSS Data Release 9 \citep{Ahn+12} or VST Atlas surveys \citep{Shanks+13, Shanks+15}.
Galaxies were assigned an integer between 0 (for ellipticals) and 3 (for late-type spirals and irregulars), with half-integers reserved for galaxies where the classification was split between two morphological types.
For 148 galaxies, $\sim$5\% of the SGS, the classification was deemed to be too uncertain, and so we do not include these galaxies when separating by visual morphology.

Stellar masses are taken from the SGS sample catalogue \citep{Bryant+15}.
These were derived from the rest-frame \textit{i}-band absolute magnitude and $g-i$ colour by using the colour-mass relation following the method of \cite{Taylor+11}.
For estimating the stellar masses, a \cite{Chabrier+03} initial mass function and exponentially declining star formation history was assumed.
Cluster galaxy masses in \cite{Owers+17} were estimated using the same method.
Measurements of the stellar velocity dispersions, $\sigma_e$, are taken from SAMI Public Data Release 3 \citep{Croom+21}.
Local surface density measurements, $\Sigma_5$, were measured from the distance to the fifth nearest neighbour within a volume-limited sample as described in \cite{Brough+17} using redshifts and photometry from GAMA \citep{Hopkins+13} and the SAMI cluster survey \citep{Owers+17}.

\section{Method}
\label{sec:method}

\subsection{Spectral fitting} 
\label{subsec:spectral_fitting}

The galaxy spectra used here are taken from the internal v0.12 aperture spectra, specifically the elliptical $r_e$ (MGE) apertures, and are equivalent to those released as part of the public Data Release 3.
We limit our selection to only those galaxies with $z\leq0.072$, due to calibration issues expanded on in Appendix \ref{app:redshift_offsets}.
To determine the stellar velocity dispersion, and correct for emission, we fit each galaxy to a set of template spectra using the Penalized Pixel-Fitting method from \cite{Cappellari+17}, \textsc{ppxf}.
For the templates, we use the empirical stellar spectra from the MILES spectral library, published by \cite{Vazdekis+11}, which are uncontaminated by emission lines. 
We use \textsc{ppxf} to fit each galaxy spectrum three times. 
We assume a Gaussian line-of-sight velocity distribution, and hence only measure the stellar velocity $v$, and stellar velocity dispersion $\sigma$.
In all cases, we include an additive 12$^{\rm{th}}$ order Legendre polynomial to account for small errors in the flux calibration, following \cite{van_de_Sande+17a}.

For the initial fit, we assume a uniform noise spectrum, such that \textsc{ppxf} weights all pixels equally, without masking out any wavelength regions.
The aperture variance spectrum allows us to estimate the standard deviation of the flux spectrum, which we take as a reliable measure of the pixel-to-pixel variation in noise.
We clip this spectrum to remove large spikes due to bad pixels, likely from cosmic rays or atmospheric emission subtraction.
If the residuals from the initial fit ($x_i$) are randomly distributed as a Gaussian, and the noise spectrum ($\sigma_i$) is correctly scaled, then ${x_i}/{\sigma_i}$ should be distributed as a standard Gaussian, with a mean of zero, and unit width.
Since this is not the case for the majority of galaxies investigated, we enforce a scaling factor on the noise spectrum. 
We take this as the ratio of the median values of the fit residuals to the noise spectrum, such that the new error spectrum can be expressed as $\sigma_{new} = \sigma_{old}\left({{\rm{med}}(x)}/{{\rm{med}}(\sigma_{old})}\right)$.
Values for this scaling were typically in the region of $\sigma_{new}\sim0.9\,\sigma_{old}$.

Following this, we perform a second fit to the data using the scaled noise spectrum.
We now make use of the \textsc{clean} keyword, which employs a 3-sigma-clipping method during the fit. 
This rejects anomalous pixels in the spectrum, such as those contaminated by emission-line infill, or bad pixels on the detector. 
We expand these regions of anomalous pixels by $\pm$25\%, in order to account for traces of emission that may not have been picked up. 
We then conduct a final fit with the scaled noise spectrum, excluding the set of anomalous pixels we have just flagged. 
This gives us a measurement for $v$ and $\sigma$, and the \textsc{ppxf} best-fit output, which is a linear combination of the best-fitting template spectra, and the 12$^{\rm{th}}$ order additive polynomial.

Errors are estimated following the bootstrap method of \cite{van_de_Sande+17a}.
We divide the spectrum into 10 regions, then randomly reallocate the residuals from the final fit within each regions.
We add these residuals to the input galaxy spectrum, then re-measure $v$ and $\sigma$ using the best fit template.
After repeating this process 100 times, we take the standard deviations of these measurements as the errors on $v$ and $\sigma$.

Finally, we replace anomalous pixels in the input spectrum with the corresponding pixels in the best-fit \textsc{ppxf} output, resulting in a galaxy spectrum corrected for all but the weakest emission-line infill and other contamination.
For any pixels that are changed, we also replace the corresponding pixels in the noise spectrum with the median of the noise over the remaining good pixels, so that they do not bias any fits in future calculations.

We note there likely exists some minute systematic effects from the emission line infill corrections.
In particular, this will affect the Balmer indices, which in turn may influence our determination of $[\alpha/\text{Fe}]$.
However, for lower S/N galaxies, we find the correction method used gives more reliable index measurements than subtracting best-fit Gaussians, in agreement with \citep{Scott+17}, and therefore this method to be a suitable compromise.

\subsection{Index measurements} \label{sec:index_measurements}

To measure the line indices, we begin by shifting the observed spectra to the restframe, based on the measured stellar recession velocities from \textsc{ppxf}. 
Since the spectral resolution of the original Lick/IDS system varies with wavelength, we must transform our spectra to match this, using the values of \cite{Worthey+97}.
We utilise the method first developed by \cite{Graves+08}, and implemented as part of their \textsc{ez\_ages} package.
We convolve our observed spectra with a wavelength-dependent Gaussian, with standard deviation $\sigma_{vel}^2+\sigma_{inst}^2+\sigma_{app}^2 = \sigma_{IDS}^2$, where $\sigma_{vel}$ is the intrinsic broadening due to the velocity dispersion of the galaxy, $\sigma_{inst}$ is the instrumental broadening, $\sigma_{IDS}$ is the Lick/IDS system resolution, and $\sigma_{app}$ is the additional broadening we apply to match the Lick/IDS resolution.
For each absorption feature, we cut a large window out of the spectrum, including the side bandpasses, and apply this broadening individually. 
Since our spectra and SSP models are both flux-calibrated, we do not also transform to the IDS response curve.

We utilise a modified version of the method developed by \cite{Cenarro+01} to measure the line indices, detailed thoroughly in Appendix \ref{app:index_measurements}.
We estimate the uncertainties on all index measurements through a Monte Carlo method.
Gaussian-distributed random noise is added to 100 realisations of the flux spectrum, using the scaled error spectrum from the \textsc{ppxf} template fitting as the standard deviation of the noise for each pixel.
The indices are re-measured for each realisation, and the standard deviation of these values is taken as the uncertainty on the initial index measurement.
Indices where fewer than 75\% of pixels in the central feature were classified as good during the \textsc{ppxf} fit are not measured, and were rejected from any further calculations.

\subsection{Index corrections} 
\label{subsec:index_corrections}

For some galaxies and indices, the combination of the intrinsic and instrumental broadening is already greater than the required Lick/IDS resolution.
Since a bias in the index measurements can lead to significant errors in the inferred stellar population parameters, we correct the measured indices following a similar method to \cite{Schiavon+07}.
To estimate the required corrections, we make use of the MILES SSP spectral models from \cite{Vazdekis+15}, which span a wide range of ages, metallicities, and abundances.
We broaden the spectra to the Lick/IDS resolution, and measure all indices, following the same method used for the SGS galaxies in \ref{sec:index_measurements}.
We then convolve each spectrum to a fixed velocity dispersion, in steps of 5 km\,s$^{-1}$, over the range 180-500 km\,s$^{-1}$, and re-measure the Lick indices.
The lower end of this range is derived from the minimum galaxy velocity dispersion which would require index corrections, and the upper end chosen to encompass all galaxies in the SGS.
For indices defined as atomic (see Appendix \ref{app:index_measurements}), we derive a multiplicative correction factor from the difference between the broadened and original spectral indices, and an additive correction for molecular indices.
These factors are collapsed over the full range of metallicities spanned by the MILES SSP models, and into 4 discrete age bands of 1.5 Gyr, 3.5 Gyr, 8 Gyr, and 14 Gyr.
Therefore, for each measured index where the intrinsic and instrumental broadening exceeds the IDS resolution, we can apply an iterative correction, where we check to see if the returned SSP age converges to the range covered by the correction factor.

\subsection{Stellar population parameters} \label{sec:stellar_population_parameters}

Having measured the Lick indices, we need to find the corresponding SSP parameters -- age, metallicity, and $\alpha$-element abundance.
We primarily utilise the SSP models of \cite{TMJ+11}, which predict the Lick index measurements as a function of $\log_{10}(\textrm{age})$, $\textrm{[Z/H]}$, and $[\alpha/\textrm{Fe}]$, spaced over a regular grid.
For each index, we interpolate these predictions onto a finer mesh, with a spacing of 0.02 in $\log_{10}(\textrm{age})$ and $\textrm{[Z/H]}$, and 0.01 in $[\alpha/\textrm{Fe}]$.
The best-fitting solution is then found by following the $\chi^2$ minimisation process of \cite{Proctor+04}.
For each point in the SSP parameter space, we calculate the reduced $\chi^2$ statistic,
\begin{equation} \label{eq:chi_squared}
    \chi^2_{\nu} = \frac{1}{\nu} \sum_{i=1}^{n} \frac{ \left( O_i - E_i \right)^2 }{\sigma_i^2},
\end{equation}
where $O_i$ is the observed index value with uncertainty $\sigma_i$, $E_i$ is the model prediction, and we sum over $n$ indices with $\nu$ degrees of freedom.
The SSP parameters that best reproduce the measured indices are then derived by finding the minimum in $\chi^2_{\nu}$ space.
The errors on these parameters can simply be read off from the $(\chi^2_{\nu}+1)$ contour, giving us the 1$\sigma$ uncertainty.
In our implementation, for each SSP parameter the 3D $\chi^2_{\nu}$ space is collapsed along the other two parameters, ensuring that the exact extent of the $(\chi^2_{\nu}+1)$ surface is found, and thus preventing an underestimation of the uncertainties.

For each galaxy, we attempt to utilise all 20 measured indices. 
Indices are rejected from the fit if they lie more than $1\sigma$ beyond the extent of the model, so as not to bias the solution.
No result is returned if fewer than five indices are available, or if the indices do not include at least one Balmer index, and one Fe index.
Whilst the line index method overemphasises the importance of Magnesium in our estimation of $[\alpha/\text{Fe}]$, we note that the contribution of other elements such as Carbon make this more than just a measure of $[\text{Mg}/\text{Fe}]$.
We also draw attention to a small caveat with the models of \cite{Thomas+10}.
At low metallicities, the Balmer index measurements for nine galaxies lie outside the SSP model grid, and so the SSP-equivalent ages are unreliable in this regime \citep{Kuntschner+10}.
This may affect our measurements for these galaxies, since our determination of $[\alpha/\text{Fe}]$ is not completely independent of age, although we do not consider this to be a significant influence, bearing in mind the very small number of galaxies affected.

\subsection{Line fitting} 
\label{subsec:line_fitting}

For all linear fits, we make use of the Python library \textsc{lmfit} by \cite{lmfit}, where we minimise the quantity
\begin{equation} \label{eq:lmfit_chi}
    \chi^2 = \sum^N_{j=1} \frac{\left[a(x_j-x_0)+b-y_j\right]^2}{
    (a\Delta x_j)^2 + (\Delta y_j)^2},
\end{equation}
adopted from \cite{Tremaine+02}.
We set $x_0$ to zero unless otherwise stated, to simplify comparisons throughout the paper and with other studies.

\subsection{Quality cuts}
\label{sec:quality_cuts}

\begin{figure}
    \centering
    \includegraphics[width=\columnwidth]{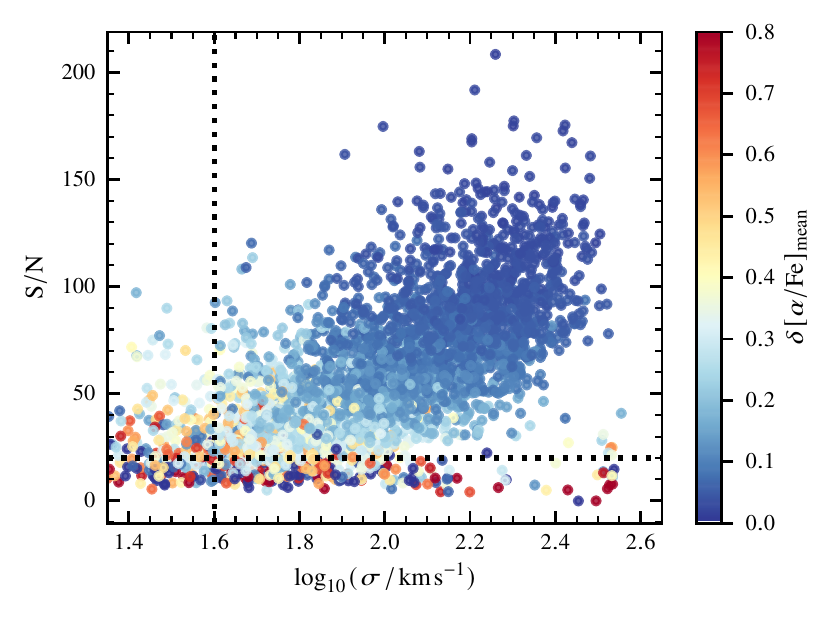}
    \caption[]{The distribution of the SGS $r_e$ aperture spectra in the velocity dispersion-S/N plane, coloured by the mean uncertainties on $[\alpha/\text{Fe}]$. The plot has been truncated at $\text{log}_{10}(\sigma)=1.35$ for clarity. We also note the strong correlation between $\sigma$ and S/N for our sample, and display the cuts made via the dotted lines.
}
    \label{fig:SN_sigma_map}
\end{figure}

We make a series of cuts to our sample, based on the reliability of our measurements of $[\alpha/\text{Fe}]$.
Figure \ref{fig:SN_sigma_map} shows the positions of the full sample in the $\sigma$-S/N plane, where we have measured the S/N between 4600-4800\AA\ on the original $r_e$ aperture spectra.
We reject spectra with a S/N$<20$, as this is not sufficient to adequately constrain $[\alpha/\text{Fe}]$, as demonstrated in previous studies such as \cite{Gallazzi+05}.
We also consider the dependence on $\sigma$.
If a galaxy has a sufficiently high uncertainty on $\sigma$, this should increase the uncertainty on our index measurements, since the convolution to the Lick/IDS resolution depends on the measured velocity dispersion.
In general, we have not considered this effect for our analysis, since the errors on most measurements of $\sigma$ are negligible compared to the errors on the indices themselves.
However, since the velocity dispersion uncertainties are considerably higher at low $\sigma$, due to the correlation with S/N visible in Fig.~\ref{fig:SN_sigma_map}, we find it prudent to also introduce a cut here, at $\text{log}_{10}(\sigma)=1.6$.

\subsection{Parameter space limitations}
\label{sec:parameter_space_limitations}

\begin{figure}
    \centering
    \includegraphics[width=\columnwidth]{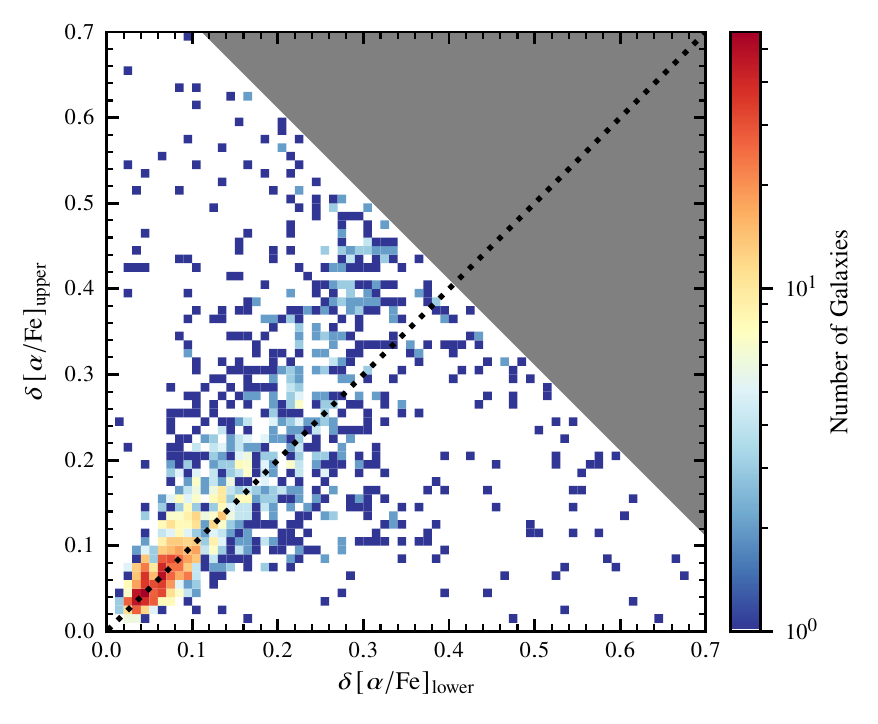}
    \caption[]{The distribution of the upper and lower 1$\sigma$ uncertainties on $[\alpha/\text{Fe}]$. 
    Points are coloured according to the number of galaxies in each bin, due to the high degree of overplotting in the region where $\delta[\alpha/\text{Fe}]<0.2$. 
    The dashed line indicates the ideal scenario, where the uncertainties for any given galaxy are equal, and the distribution is symmetrical. 
    The grey shaded region shows the extent of the model parameter space.
    Galaxies where both uncertainties are clipped by the parameter space have already been cut from the sample.  
    }    
    \label{fig:err_asymmetry}
\end{figure}
A considerable complication with the analysis is taking into account the finite limits of the SSP models. 
We exclude any galaxy in which the combined $1\sigma$ uncertainty covers the full extent of the $[\alpha/\text{Fe}]$ parameter space.
In Fig.~\ref{fig:err_asymmetry}, we plot the upper and lower $1\sigma$ uncertainties on $[\alpha/\text{Fe}]$ against each other for the entire sample.
In an ideal world, all galaxies would lie exactly on the dashed line, since we do not expect any asymmetry in the errors.
However, beyond a small degree of scatter, we find that for galaxies where $\delta[\alpha/\text{Fe}]_{\rm{upper}}>0.3$, there is no corresponding increase in the lower uncertainty bound.
After investigating the underlying correlations, we find that these galaxies ($\sim$10\% of the sample) are typically low-$\sigma$, with low $\alpha$ abundance. 
This low $\alpha$ abundance combined with a high measurement error leads to $\delta[\alpha/\text{Fe}]_{\rm{lower}}$ being clipped by the edges of the parameter space. 
When performing weighted fits using Equation \ref{eq:lmfit_chi}, these galaxies may unfairly bias the solution.
For the remaining galaxies unaffected by this clipping, we calculate the asymmetry of the uncertainties, $\delta[\alpha/\text{Fe}]_{\rm{upper}}-\delta[\alpha/\text{Fe}]_{\rm{lower}}$.
We find the mean of this distribution to be 0.017\,dex, with a standard deviation of 0.052\,dex.
Although this would indicate a slight systematic offset, once we take into account the resolution limit of 0.01 in $[\alpha/\text{Fe}]$, we conclude that the asymmetry is not significant.
Therefore, for the galaxies where one of the $1\sigma$ limits is bounded by the model range, we force the uncertainties to be symmetrical, taking the unbounded uncertainty as the guide. 

\subsection{Completeness} \label{subsec:completeness}

\begin{table*}
    \centering
    \caption[]{
    The total number of galaxies used for each stage of the analysis, separated by both optical and kinematic morphology. 
    Intermediate classifications have been grouped with the earlier of the two types, e.g. E/S0 galaxies contribute to the total under the E column.
    For analyses of ETGs, we use only galaxies classified as E, E/S0, and S0.
    Note that the total number of galaxies differs slightly between sections, since the required properties cannot be measured for all galaxies.
    }
    \label{tab:total_number_of_galaxies}
    \begin{tabular}{rccccccl}\hline\hline
        \multirow{2}{*}{Sample} & \multirow{2}{*}{Total} & \multicolumn{5}{c}{Optical Morphology} & \multirow{2}{*}{Section} \\\cline{3-7}
        & & E & S0 & Sa/b & Sc & Unclassified &\\\hline
        All Observations & 3426 & 603 & 785 & 682 & 1182 & 174 &\\
        Unique Galaxies & 3068 & 561 & 728 & 605 & 1026 & 148 &\\
        $z\leq0.072$ & 2773 & 459 & 630 & 544 & 1002 & 138 & \ref{app:redshift_offsets} \\
        Quality Cuts & 2093 & 453 & 623 & 524 & 404 & 89 &  \ref{sec:quality_cuts} \\
        $\rm{log}_{10}\left(\Sigma_5\right)\leq1.15$ & 1416 & 216 & 363 & 424 & 362 & 51  & \ref{sec:environmental_dependence} \\
        $\rm{log}_{10}\left(\Sigma_5\right)>1.15$ & 660 & 230 & 252 & 95 & 46 & 37 & \ref{sec:environmental_dependence} \\
        \multirow{2}{*}{} & \multirow{2}{*}{} & \multicolumn{5}{c}{Early Type Galaxies} &\\\cline{3-7}
        & & \multicolumn{2}{c}{E} & \multicolumn{2}{c}{E/S0} & S0 &\\
        $\rm{log}_{10}\left(\Sigma_5\right)\leq1.15$ & 379 & \multicolumn{2}{c}{111} & \multicolumn{2}{c}{104} & 164 & \ref{sec:environmental_dependence} \\
        $\rm{log}_{10}\left(\Sigma_5\right)>1.15$ & 375 & \multicolumn{2}{c}{130} & \multicolumn{2}{c}{99} & 146 & \ref{sec:environmental_dependence} \\
    \end{tabular}
\end{table*}

\begin{figure}
    \centering
    \includegraphics[width=\columnwidth]{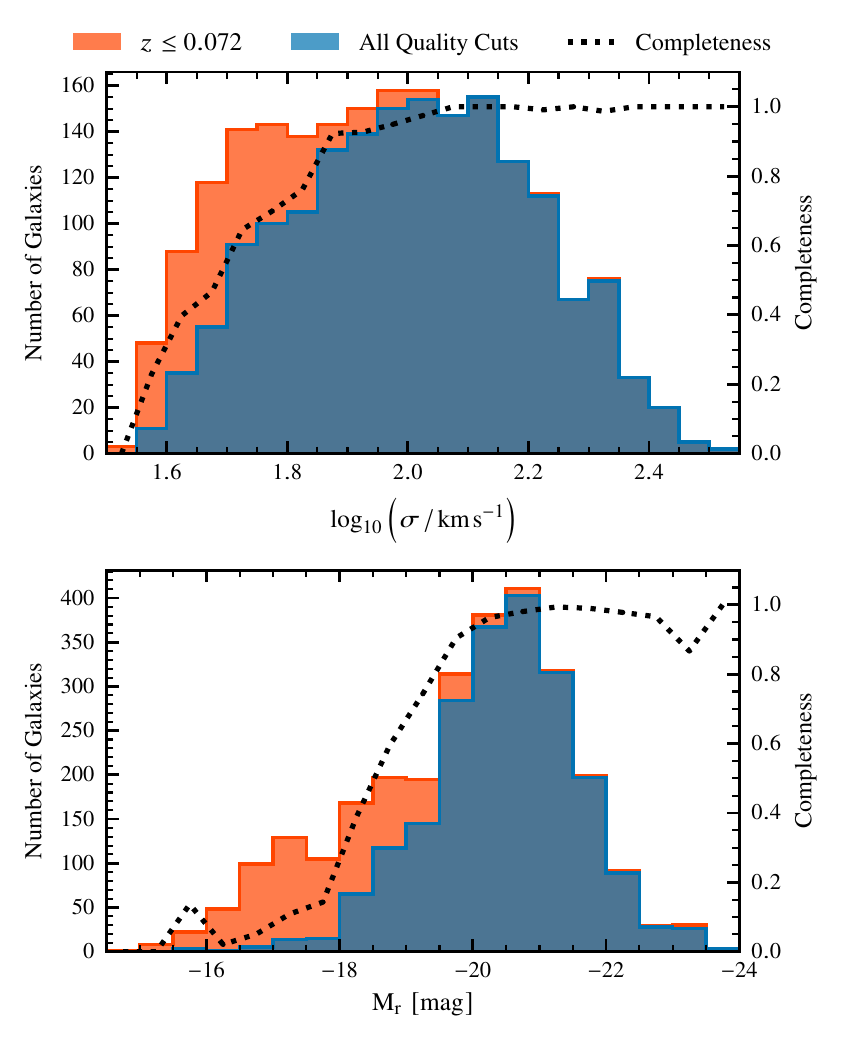}
    \caption[]{The number of unique galaxies in both the redshift-limited sample, and the core sample after all quality cuts, consisting of 2773 and 2093 galaxies respectively.
    We plot the distribution of these galaxies as a function of $\rm{log}_{10}\left(\sigma\right)$, and absolute magnitude in the $r$-band, $M_r$.
    We overlay the completeness of our final sample after all quality cuts, relative to the number of unique galaxies with $z\leq0.072$.}
    \label{fig:quality_cut_hist}
\end{figure}

In total, 975 unique galaxies were removed from the sample due to the cumulative quality cuts outlined in this section.
We rejected 239 galaxies with $\rm{log}_{10}\left(\sigma\right)<1.6$, and 242 with a spectral $\rm{S/N}<20$.
199 galaxies either fall on the outer limits of the parameter space in $[\alpha/\text{Fe}]$ or have uncertainties spanning the entire range.
Finally, we removed 295 galaxies with $z>0.072$ due to possible skyline contamination.
We advise a degree of caution here, since the exact numbers will vary significantly depending on the order of the cuts.
For example, galaxies rejected with $\rm{log}_{10}\left(\sigma\right)<1.6$ could also have been rejected in the first instance for having a $\rm{S/N}<20$, as shown in Fig.~\ref{fig:SN_sigma_map}.
The number of galaxies of each type remaining are summarised in Table \ref{tab:total_number_of_galaxies}.

We also note that the total number of galaxies differs between sections of the paper, due to the limitations of the prerequisite measurements.
Quantities such as the local surface density $\Sigma_5$, cannot be measured for all galaxies, for a variety of reasons expanded upon in the relevant papers.
As such, not all of the 2093 galaxies that remain are utilised in each element of our analysis, with the decomposition tabulated in Table \ref{tab:total_number_of_galaxies}.

Figure \ref{fig:quality_cut_hist} displays how the cumulative quality cuts have affected the completeness of the sample.
We measure the completeness of our final 2093 galaxies against the 2773 galaxies with $z\leq0.072$.
Visually, it is clear that the completeness of the sample suffers at low $\sigma$, and reaches 50\% above $\rm{log}_{10}\left(\sigma\right)\sim1.7$.
Above $\rm{log}_{10}\left(\sigma\right)=2$, the sample is almost complete relative to galaxies below $z=0.072$, not falling below $90\%$.
Similarly, looking at the magnitude distribution, we find $>80\%$ of galaxies fainter than $M_r=-18$ are cut, although nearly all galaxies brighter than $M_r=-20$ remain in the sample.

Looking at Table \ref{tab:total_number_of_galaxies}, we can see the effect of the quality cuts on the sample morphology distribution.
These appear to almost exclusively affect those galaxies classified as Sc, although this is not entirely surprising given that we have introduced a cut-off in velocity dispersion.

\section{Environmental Dependence} \label{sec:environmental_dependence}

\subsection{Separation by colour} \label{sec:separation_by_colour}

\begin{figure}
    \centering
    \includegraphics[width=\columnwidth]{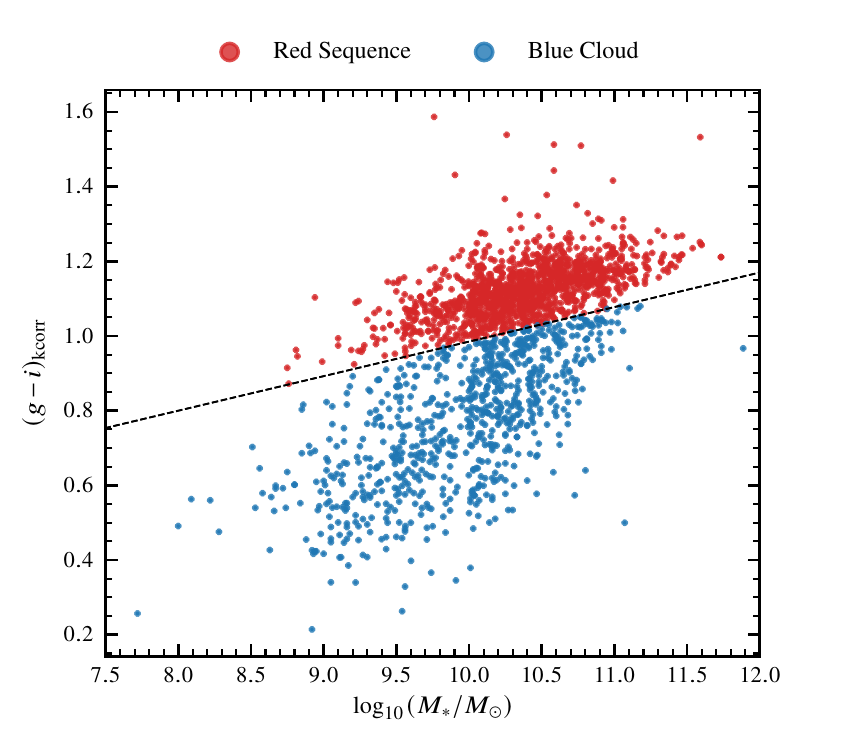}
    \caption[]{
    The distribution of $k$-corrected $(g-i)$ galaxy colours as a function of stellar mass, for our final sample.
    The dividing line between the red sequence and blue cloud is taken from \cite{Owers+17}.
    }
    \label{fig:red_sequence_selection}
\end{figure}

\begin{figure}
    \centering
    \includegraphics[width=\columnwidth]{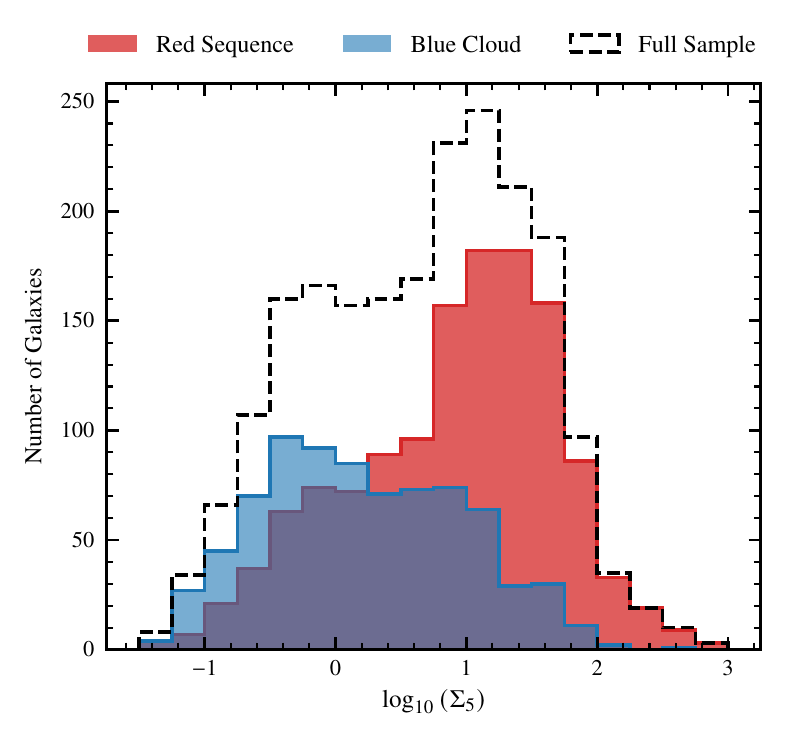}
    \caption[]{The distribution of $\rm{log}_{10}(\Sigma_5)$ for red sequence and blue cloud galaxies, alongside the full sample.
    At the highest values of $\rm{log}_{10}(\Sigma_5)$, galaxies almost exclusively belong to the red sequence, whereas we find a more balanced distribution towards lower values of $\rm{log}_{10}(\Sigma_5)$.
    }
    \label{fig:surf_dens_hist_outline}
\end{figure}

\begin{figure*}
    \centering
    \includegraphics[width=0.66\textwidth]{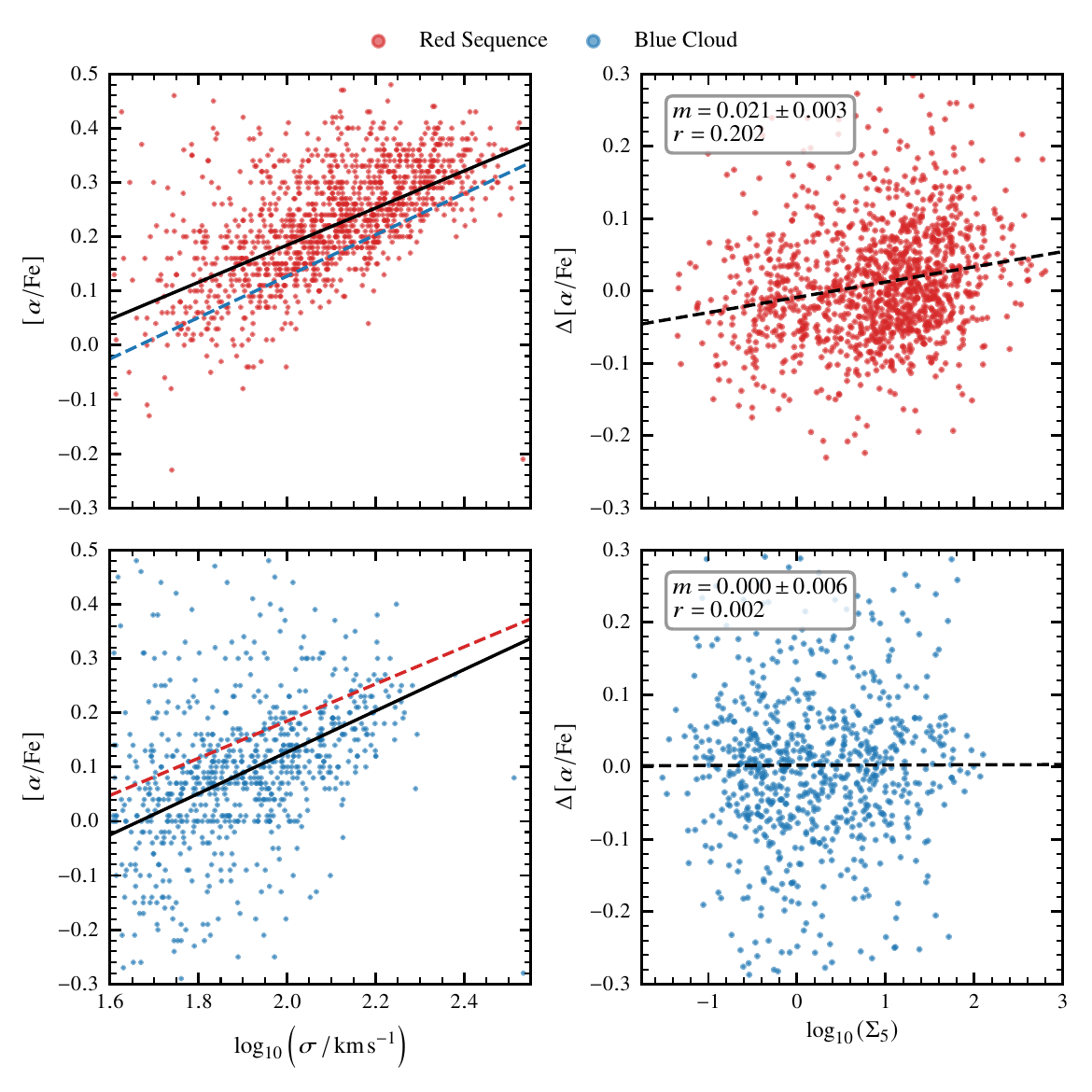}
    \caption[]{
    The left-hand figures display the $[\alpha/\text{Fe}]$-$\sigma$ relation for both the blue cloud and red sequence, with separate best fit relationships shown as solid black lines.
    For each sample, we overlay the linear relationship for the other sample as a dashed line, to allow a direct comparison.
    We then fit a second linear trendline to the residuals $\Delta[\alpha/\text{Fe}]$, expressed as a function of the local surface density $\Sigma_5$, and displayed as a dashed line in the figures on the right.
    The gradient $m$ and the correlation coefficient $r$ are inset in the relevant plots.
    The extent of the $\Delta[\alpha/\text{Fe}]$ axis has been truncated to display $\sim\,$99\% of galaxies, whilst allowing for a clear visual comparison between the samples.
    }
    \label{fig:red_blue_sigma_5_residuals}
\end{figure*}

Several previous studies, among them \cite{Baldry+06}, have found that the most significant change with environment is the fraction of red galaxies, $f_r$.
To consider the dependence of the $[\alpha/\text{Fe}]$-$\sigma$ relation on the environment, we therefore control for this variable by isolating samples of both red and blue galaxies, and separately analysing their stellar populations.
We begin by dividing our full sample into a red sequence and blue cloud, shown in Fig.~\ref{fig:red_sequence_selection}, and composed of 1292 and 776 galaxies respectively.
We utilise the $k$-corrected $(g-i)$ colour, $(g-i)_{kcorr}$, where the $k$-corrections have been determined using \textsc{calc\_kcor} code from \cite{Chilingarian+10}.
The division between the red sequence and blue cloud is taken from \cite{Owers+17}.
Figure \ref{fig:surf_dens_hist_outline} shows the corresponding distribution of galaxies in $\rm{log}_{10}(\Sigma_5)$, our primary measure of the environmental density taken from \cite{Brough+17}.
Whilst both samples are well represented at the lowest values of $\rm{log}_{10}(\Sigma_5)$, above  $\rm{log}_{10}(\Sigma_5)\sim1$ the SGS is dominated by red sequence galaxies.

We fit separate linear relations between $[\alpha/\text{Fe}]$ and $\rm{log}_{10}(\sigma)$ for both the red sequence and blue cloud, shown in Fig.~\ref{fig:red_blue_sigma_5_residuals}.
For the red sequence, we derive  $$[\alpha/\text{Fe}]=(0.342\pm0.011)\rm{log}_{10}\left(\sigma\right)-(0.50\pm0.03),$$ and for the blue cloud, $$[\alpha/\text{Fe}]=(0.381\pm0.017)\rm{log}_{10}\left(\sigma\right)-(0.63\pm0.04).$$
From this and Fig.~\ref{fig:red_blue_sigma_5_residuals}, we deduce that galaxies in the red sequence are systematically $\alpha$-enhanced over those in the blue cloud.
The steeper relationship with $\sigma$ for blue galaxies means that this effect manifests predominantly at low velocity dispersion, with an offset of $\sim$0.07\,dex in $[\alpha/\text{Fe}]$ at $\rm{log}_{10}(\sigma)=1.6$ compared to $\sim$0.04 at $\rm{log}_{10}(\sigma)=2.3$ (the velocity dispersion limit of the blue cloud).

To compare the environmental influence for each sample, we express the residuals $\Delta[\alpha/\text{Fe}]$ as a function of $\Sigma_5$, and refit a linear relation, shown in Fig.~\ref{fig:red_blue_sigma_5_residuals}.
For galaxies in the blue cloud, we find no correlation, $r\sim0$, and a very large degree of scatter in the residuals.
For red sequence galaxies, the scatter in the residuals is reduced, although still substantial.
We find a positive correlation for the red sequence, with $r\sim0.2$, which we derive as $$\Delta[\alpha/\text{Fe}]=(0.021\pm0.003)\rm{log}_{10}(\Sigma_5)-(0.009\pm0.003).$$
Red sequence galaxies in high density environments (HDEs) have higher $[\alpha/\text{Fe}]$ than their counterparts in low density environments (LDEs), at a fixed velocity dispersion.

\subsection{ETG residuals} \label{sec:etg_residuals}

\begin{figure*}
    \centering
    \includegraphics[width=\textwidth, trim=0 0 0 0.8cm, clip]{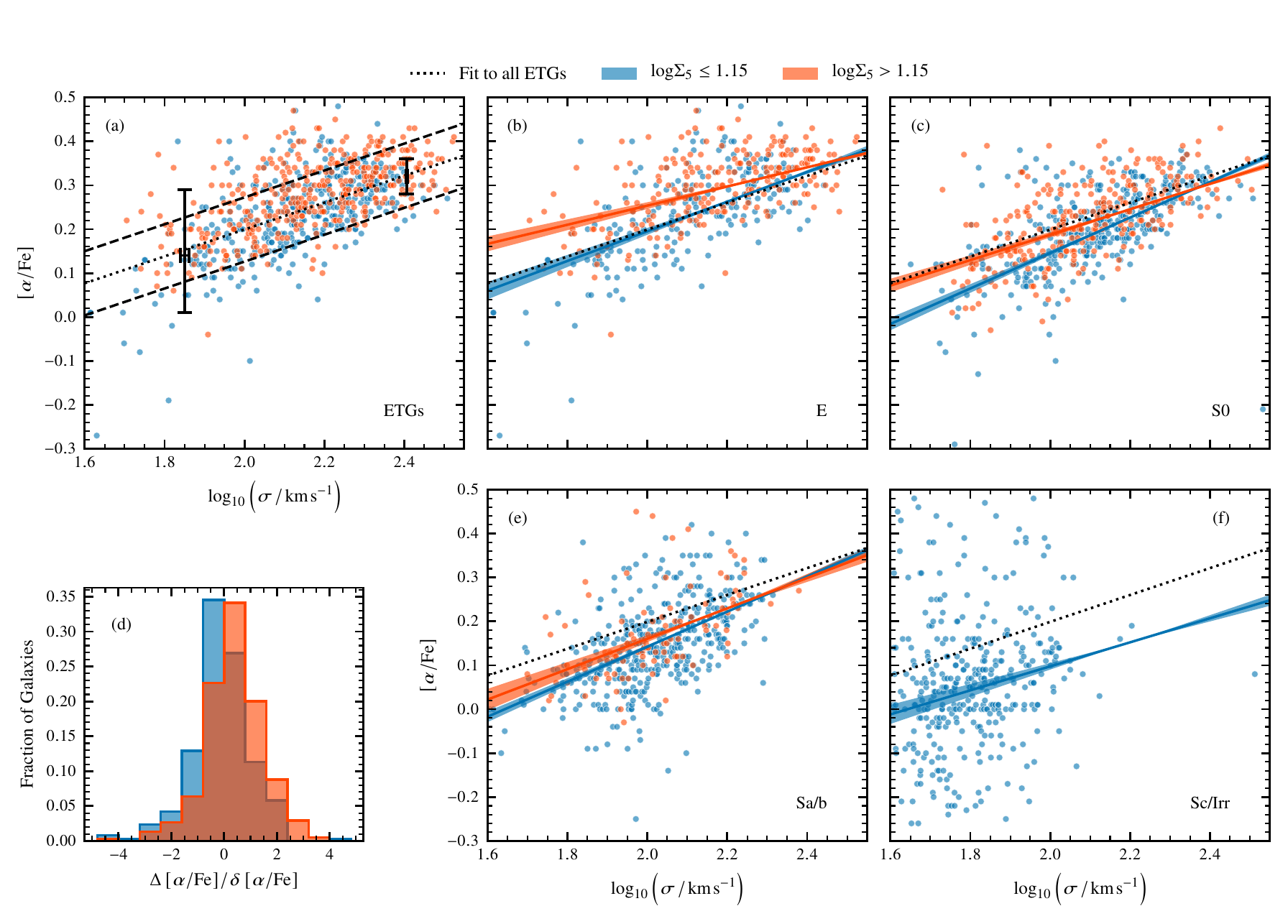}
    \caption[]{
    (a) We fit a linear trendline to our combined ETG sample, shown as a dotted line.
    The errorbars show the median uncertainties on $[\alpha/\text{Fe}]$ and $\sigma$ for the 5$^{\text{th}}$ and  95$^{\text{th}}$ percentile of ETGs ordered by velocity dispersion, and the dashed lines the $\pm1$ standard deviation of galaxies from the best fit line.
    Galaxies are coloured according to their local surface density, and we display the normalised residuals $\Delta\left[\alpha/\rm{Fe}\right]/\delta\left[\alpha/\rm{Fe}\right]$ for the ETGs in (d).
    Using the full sample of galaxies, we also fit each environment and morphology separately, without constraining the fit in any way, and show the results in (b), (c), (e), and (f).
    Since we do not consider the result for Sc galaxies with high local densities to be as reliable as the other groups due to the large difference in sample size, we do not display it here.
    We use the relationship for ETGs as a point of reference, represented by the dotted black line.
    }
    \label{fig:ETG_unconstrained_combined}
\end{figure*}

For ease of comparison with other studies, we investigate a sample of ETGs, those classified as E, E/S0, and S0 by \cite{Cortese+16}.
This sample, containing 767 galaxies, can generally be considered a subset of the red sequence discussed in Section \ref{sec:separation_by_colour}, with only 3\% of ETGs being drawn from the blue cloud.
As such, the relations derived for the red sequence hold true for ETGs.
We fit the whole ETG population to a single linear relationship, shown in Fig.~\ref{fig:ETG_unconstrained_combined}, and calculated as $$[\alpha/\text{Fe}]=(0.306\pm0.014)\rm{log}_{10}\left(\sigma\right)-(0.41\pm0.03).$$
From this, we measure the absolute residuals in $[\alpha/\text{Fe}]$, and as before, derive the relation $$\Delta[\alpha/\text{Fe}]=(0.023\pm0.003)\rm{log}_{10}(\Sigma_5)-(0.013\pm0.004),$$ with a correlation coefficient $r=0.25$.
Although this indicates a more positive correlation than the red sequence alone, we investigate alternative methods to test the statistical significance of our findings.

We divide the ETGs into two samples of approximately equal size, based on the local environmental surface density of each galaxy, giving  high, $\text{log}_{10}(\Sigma_5)>{1.15}$, and low, $\text{log}_{10}(\Sigma_5)\leq{1.15}$, environmental density samples.
We normalise the residuals $\Delta[\alpha/\text{Fe}]$ by the associated error, $\delta[\alpha/\text{Fe}]$, and show the residual distributions in Fig.~\ref{fig:ETG_unconstrained_combined}.
Visually, we find that the distribution of galaxies with high local surface densities is shifted to slightly higher values of $[\alpha/\text{Fe}]$, compared to galaxies in lower density environments, with the median value of $[\alpha/\text{Fe}]$ increasing by 0.025\,dex.

Conducting a Kolmogorov-Smirnov test on the samples allows us to reject the notion that these two samples are drawn from the same parent population ($p<0.01$), and a Mann-Whitney test similarly concludes that the medians are statistically different ($p<0.01$).
However, since the actual difference in the sample medians is small, on the order of 0.4 standard errors, this is only a small offset compared to the scatter visible in the overall relationship.
Based on this, we cannot rule out the primary cause of this offset being the well known morphology-density relation of \cite{Dressler+80}.
Indeed, in Table \ref{tab:total_number_of_galaxies}, we show the breakdown of the constituent morphologies in the ETG sample.
The fraction of galaxies labelled as E/S0 remains almost constant across the two environments.
By comparison, the fraction of ellipticals increases from 29.3\% in LDEs to 34.6\% in HDEs.
As such, the correlations between $\alpha$-enhancement and morphology shown in the following section could be sufficient to explain this difference in distributions.

\subsection{Morphology dependence} \label{sec:morphology_dependence}

Satisfied that our division by environmental density is sufficient to show a divergence in the population, without forcing the residuals to a fit against $\Sigma_5$, we further examine the environmental offset by simultaneously testing the underlying dependence on morphology.
We separate all 2,101 galaxies in the full sample by visual morphology, leaving 4 similarly sized groups.
For galaxies where the classification is split between two morphologies, we group these with the earlier type of the two, such that an E/S0 galaxy would be included in the E sample.
We use the same cutoff value for environmental density as before, $\rm{log}_{10}(\Sigma_5)=1.15$, for consistency across comparisons, and discuss this further in Appendix \ref{app:sample_selection_bias}.
The results are shown in Fig.~\ref{fig:ETG_unconstrained_combined}, with the best fit parameters in Table \ref{tab:Subplots_morph_surf_dens_unconstrained_line_fit}.

\begin{table*}
    \centering
    \caption[]{The best fit parameters for the relationships shown in Fig.~\ref{fig:ETG_unconstrained_combined}, in the form $[\alpha/\text{Fe}]=a\,\text{log}_{10}(\sigma/100)+b$, where we use the notation $b\equiv[\alpha/\text{Fe}]_{100}$. 
    Note the substantially larger uncertainties on the parameters for Sc galaxies with high local surface densities.}
    \label{tab:Subplots_morph_surf_dens_unconstrained_line_fit}
    \begin{tabular}{@{\extracolsep{8pt}}rcccc@{}}\hline\hline
        \multirow{2}{*}{Morphology} & \multicolumn{2}{c}{$\text{log}_{10}(\Sigma_5)\leq1.15$} & \multicolumn{2}{c}{$\text{log}_{10}(\Sigma_5)>1.15$} \\\cline{2-3}\cline{4-5}
        & $a$ & $[\alpha/\text{Fe}]_{100}$ & $a$ & $[\alpha/\text{Fe}]_{100}$ \\\hline
        E & $0.34\pm0.03$ & $0.195\pm0.008$ & $0.22\pm0.02$ & $0.253\pm0.007$\\
        S0 & $0.40\pm0.02$ & $0.146\pm0.005$ & $0.29\pm0.02$ & $0.188\pm0.006$\\
        Sa/b & $0.39\pm0.02$ & $0.144\pm0.004$ & $0.35\pm0.04$ & $0.160\pm0.008$\\
        Sc & $0.28\pm0.04$ & $0.099\pm0.007$ & $-0.08\pm0.13$ & $0.06\pm0.02$\\\hline   
    \end{tabular}
\end{table*}

For galaxies in LDEs, there is little change in the best fit relation with morphology, with the S0 and Sa/b groups entirely consistent within the stated errors.
The slope of the relation is slightly shallower for elliptical galaxies, with the change driven by ellipticals at low velocity dispersion, $\rm{log}_{10}(\sigma)<2$, being $\alpha$-enhanced over other morphologies.
Conversely, at high velocity dispersion, $\rm{log}_{10}(\sigma)>2$, the measured $[\alpha/\text{Fe}]$ ratio is consistent across most morphological types in LDEs.
The exception to this is the Sc/Irr group, which shows some evidence for a shallower relationship than all other morphologies.
This may be partially caused by the limited sampling, with only 6.9\% of Sc/Irr galaxies having $\rm{log}_{10}(\sigma)>2$.

For galaxies in high density environments (HDEs), the picture is more complicated.
In Table \ref{tab:Subplots_morph_surf_dens_unconstrained_line_fit}, for all morphologies earlier than Sc, in the same manner as for galaxies in LDEs, there is no observable difference between the HDE samples at high velocity dispersions.
The difference here occurs primarily at low $\sigma$, with galaxies classified as S0 being $\alpha$-enhanced over their Sa/b counterparts by approximately 0.04\,dex at $\rm{log}_{10}(\sigma)=1.6$.
This offset increases further when considering the elliptical galaxies, which are $\alpha$-enhanced by an additional 0.09\,dex over S0 galaxies at  $\rm{log}_{10}(\sigma)=1.6$.

In Sc galaxies, recent bursts of star formation will strongly influence the integrated light, making it difficult to evaluate the abundance ratio.
Thus, no conclusions can be drawn from the small sample of Sc galaxies.



Comparing now the same morphological types across environments, in the high-$\sigma$ regime, we find no significant difference between galaxies in HDEs or LDEs, irrespective of morphological type.
Hence, for galaxies with an earlier type than Sc, at high velocity dispersion there is little evidence here to support a difference in the relative abundance of $\alpha$-elements across environments.
Conversely, at low $\sigma$, there is a substantial offset between best fit relations for the low and high density samples, with galaxies in HDEs being $\alpha$-enhanced over those in LDEs.
At $\rm{log}_{10}(\sigma)=1.6$, the lower limit of our selection, we find an offset of 0.09 dex in $[\alpha/\text{Fe}]$ for elliptical galaxies, 0.08 dex for S0, and narrowing to 0.05 for the Sa/b grouping.
There therefore appear to be two components.
The first is that low-$\sigma$ galaxies residing in a similar environment show an increase in $[\alpha/\text{Fe}]$ when looking along the Hubble sequence from spirals to ellipticals.
The second is that when comparing galaxies of the same morphological type, those residing in denser environments are likely to be $\alpha$-enhanced over their LDE counterparts.

\subsection{Separation by velocity dispersion} \label{sec:separation_by_velocity_dispersion}
\begin{figure*}
    \centering
    \includegraphics[width=0.66\textwidth]{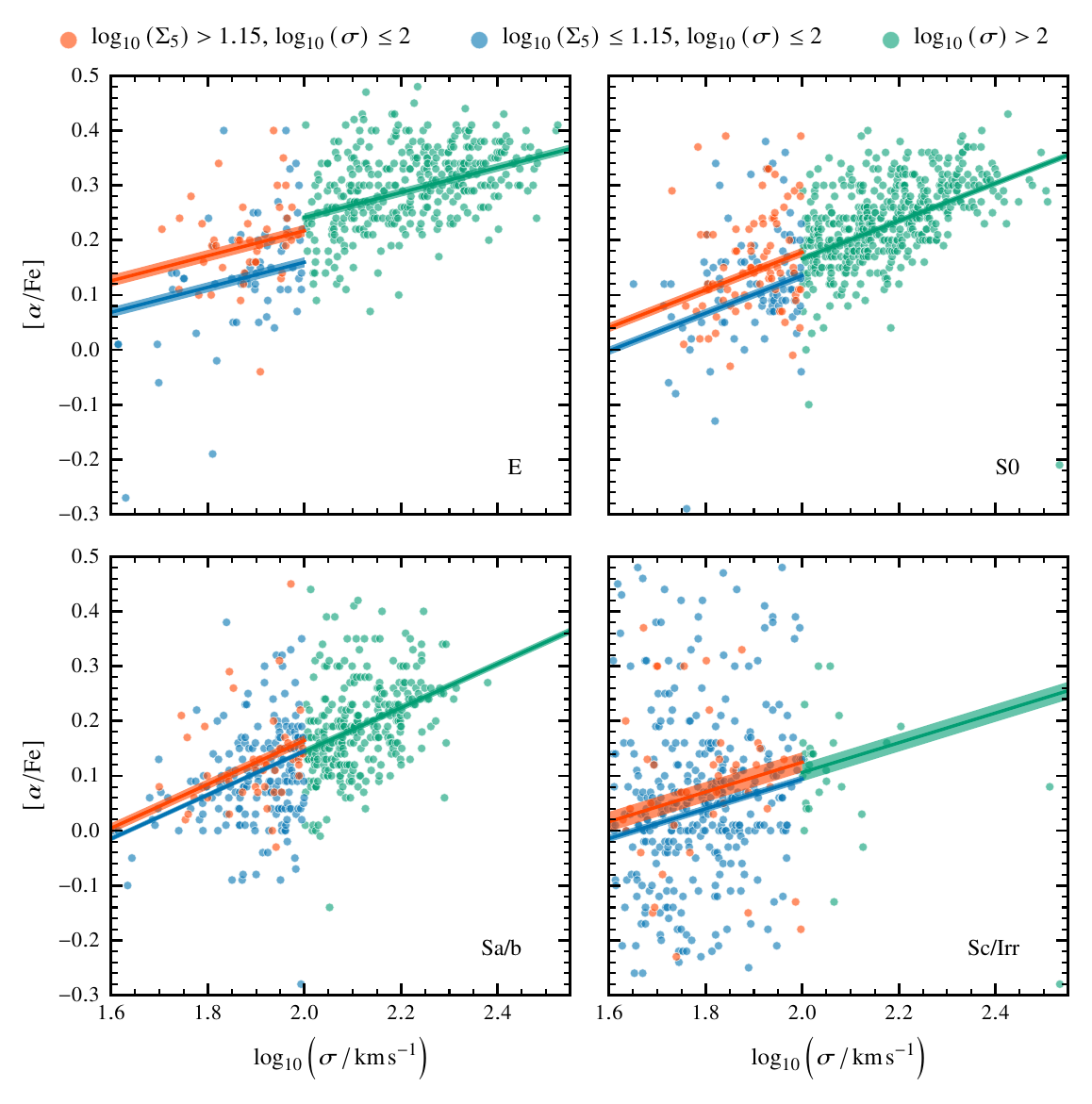}
    \caption[]{For each morphology, we fit linear trendlines to all galaxies above $\text{log}_{10}(\sigma)>2$. We then fit each environment separately for $\text{log}_{10}(\sigma)\leq2$, fixing the gradient to the value measured at higher $\sigma$.}
    \label{fig:Subplots_morph_surf_dens_sigma_cut_line_fit}
\end{figure*}

\begin{table*}
    \centering
    \caption[]{
    The best fit parameters for the relationships shown in Fig.~\ref{fig:Subplots_morph_surf_dens_sigma_cut_line_fit}, where the gradient $a$ is fixed by fitting to galaxies where $\text{log}_{10}(\sigma)>2$.
    }
    \label{tab:Subplots_morph_surf_dens_sigma_cut_line_fit}
    \begin{tabular}{@{\extracolsep{4pt}}rcccc}\hline\hline
        \multirow{2}{*}{Morphology} & \multirow{2}{*}{$a$} & $[\alpha/\text{Fe}]_{100}$ & \multicolumn{2}{c}{$[\alpha/\text{Fe}]_{100,\text{\ fixed\ }a}$} \\\cline{3-3} \cline{4-5}
        & & $\text{log}_{10}(\sigma)>2$ & $\text{log}_{10}(\Sigma_5)\leq1.15$ & $\text{log}_{10}(\Sigma_5)>1.15$ \\\hline
        E & $0.23\pm0.02$ & $0.241\pm0.008$ & $0.160\pm0.009$ & $0.217\pm0.011$\\
        S0 & $0.34\pm0.02$ & $0.166\pm0.006$ & $0.136\pm0.008$ & $0.178\pm0.008$\\
        Sa/b & $0.40\pm0.04$ & $0.144\pm0.006$ & $0.145\pm0.005$ & $0.164\pm0.009$\\
        Sc & $0.27\pm0.12$ & $0.106\pm0.016$ & $0.094\pm0.007$ & $0.124\pm0.018$\\\hline
    \end{tabular}
\end{table*}

It is worth considering however, that the offsets at low $\sigma$ discussed in the previous section could be changed substantially, depending on the exact routine used to simultaneously fit both the intercept and gradient for each subsample.
For this reason, we also utilise a similar method to \cite{Liu+16}, shown in Fig.~\ref{fig:Subplots_morph_surf_dens_sigma_cut_line_fit}.
Due to the limited differences between high-$\sigma$ galaxies across environments, for each morphology, we fit all galaxies with $\text{log}_{10}(\sigma)>2$ to a single linear relationship.
Galaxies where $\text{log}_{10}(\sigma)\leq2$ are then divided into high and low density environments as before, and the gradient for their best fit relationships is fixed to the value derived at high $\sigma$.
The resulting parameters are summarised in Table \ref{tab:Subplots_morph_surf_dens_sigma_cut_line_fit}.

One immediate result from this method, as opposed to that used in Section \ref{sec:morphology_dependence}, is that the results from the Sc sample no longer appear anomalous compared to the other morphologies, despite the small sample of galaxies with $\text{log}_{10}(\sigma)>2$.
The overall results are largely consistent with the conclusions drawn from Fig.~\ref{fig:ETG_unconstrained_combined}, with the largest offset in $[\alpha/\text{Fe}]$ occurring in elliptical galaxies.
For Sa/b/c galaxies, we find that the different intercepts for the low and high surface density samples have overlapping uncertainties.
In combination with Fig.~\ref{fig:ETG_unconstrained_combined}, we take this as evidence that there is little measurable difference between the stellar populations of late type galaxies with low and high local surface densities.
Conversely, for ETGs, we find that there is sufficient evidence for low-$\sigma$ galaxies in high density environments having a truncated period of star formation relative to their counterparts in lower density environments.

\begin{figure}
    \centering
    \includegraphics[width=\columnwidth]{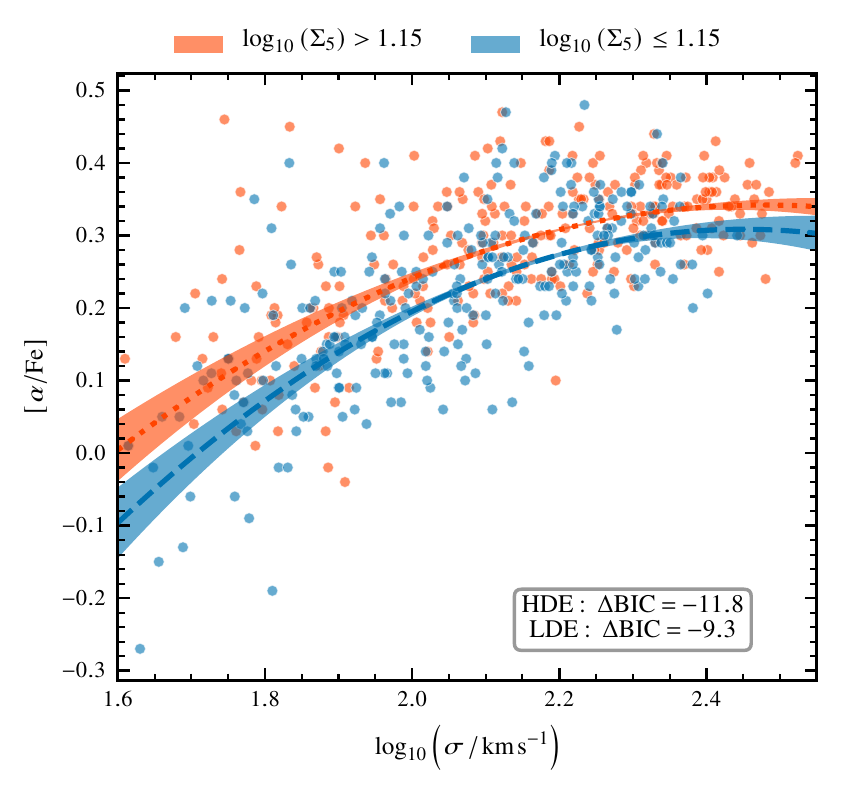}
    \caption[]{For each subsample of elliptical galaxies, we fit a parabolic relationship of the form $[\alpha/\text{Fe}]=a\,\text{log}_{10}(\sigma)^2+b\,\text{log}_{10}(\sigma)+c$.
    The change in the Bayesian Information Criteria, $\Delta\rm{BIC}$, is calculated compared to the linear relationships shown for elliptical galaxies in Fig.~\ref{fig:ETG_unconstrained_combined}}
    \label{fig:parabola}
\end{figure}

In Fig.~\ref{fig:Subplots_morph_surf_dens_sigma_cut_line_fit}, the value of $b$ for high-$\sigma$ elliptical galaxies is higher than either of the values of $b$ for low-$\sigma$ galaxies.
Investigating this, we fit to a parabolic function of the form $[\alpha/\text{Fe}]=a\,\text{log}_{10}(\sigma)^2+b\,\text{log}_{10}(\sigma)+c$, and show the results in Fig.~\ref{fig:parabola}, where we separate out galaxies in LDEs and HDEs.
To distinguish between the linear model of Fig.~\ref{fig:ETG_unconstrained_combined}, we utilise the Bayesian Information Criterion \citep[BIC,][]{Schwarz+78}.
The BIC includes a penalty term for the number of parameters to prevent overfitting, and the model with the lowest BIC is preferred.
The difference in favour of the parabolic model is substantial, with $\Delta\rm{BIC}=-11.8$ for galaxies in HDEs, and $\Delta\rm{BIC}=-9.3$ for those in LDEs.
Hence, we find strong support for a flattening of the $[\alpha/\text{Fe}]$-$\sigma$ relationship in high velocity dispersion elliptical galaxies.
We find no such support for other morphological classifications.
The choice of a linear or parabolic function does not affect our conclusions regarding $\alpha$-enhancement at low velocity dispersion, although the offset is now greater than $0.03$\,dex even at high $\sigma$, and would benefit from further study.

\section{Discussion} \label{sec:discussion}

\subsection{Comparisons to previous studies}

The multitude of methods of calculating the values of age, metallicity and $\alpha$-abundance complicate any comparisons of absolute values across different studies.
Small changes in the method, such as a different set of stellar population models, can cascade into larger systematic offsets in the final results.
Hence, we only consider here the relative trends and scaling relations.

One of the most obvious comparisons is to \citetalias{Scott+17}, since we are both working with galaxies drawn from the SGS and using a similar method to determine the stellar populations (a full comparison of the measurements themselves is contained in Appendix \ref{app:measurement_comparison}).
However, where \citetalias{Scott+17} predominantly consider the SSP-equivalent parameters as a function of mass, we instead use $\sigma$, limiting direct comparisons.
We do not explicitly compare morphologies over all environments, but our results in Fig.~\ref{fig:Subplots_morph_surf_dens_sigma_cut_line_fit} would indicate that we find a similar offset in $[\alpha/\text{Fe}]$ ranging from E to Sc galaxies, on the order of $0.066\pm0.011$\,dex in LDEs, and $0.093\pm0.021$\,dex in HDEs.
In contrast to \citetalias{Scott+17}, we find the morphology dependence occurring predominantly at low $\sigma$, and similarly for our environmental dependence, which we estimate at $\sim0.06$\,dex, for galaxies with $\rm{log}_{10}(\sigma)<2$.
This may simply be a function of our distinct sample selection, since we have simultaneously separated the dependence on morphology and environment (an option which was not available to \citetalias{Scott+17} at the time, as the SGS was still in progress).

Another study using the SGS is that of \cite{Pak+21}, which investigated passive spirals, using S0 galaxies as a control sample.
We reach markedly different conclusions for the S0 group; where we find a divergence in $[\alpha/\text{Fe}]$ at low $\sigma$, Pak et al. found an offset occurring predominantly at $\sigma\gtrsim100\rm{km\,s}^{-1}$.
We note that Pak et al. retained field galaxies with the highest recession velocities, which we show in Appendix \ref{app:redshift_offsets} to have an offset in $[\alpha/\text{Fe}]$, and believe to be anomalous.
Since these are also high velocity dispersion galaxies, we are confident that this hypothesis accounts for the erroneous conclusions.

Comparing to the ETG sample of \citetalias{McDermid+15}, our results in Fig.~\ref{fig:ETG_unconstrained_combined} are in excellent agreement.
For the derived best fit relations, we find both the gradients and intercepts are consistent within the stated errors.
Similarly, we fit our ETG sample to a single relationship, and separate the residuals $\Delta[\alpha/\text{Fe}]$ by environmental density.
Performing a KS test on these residuals, our results also show a very high probability of the HDE and LDE samples being drawn from distinct parent distributions, in good agreement with \citetalias{McDermid+15}.
Both our samples cover a consistent range in velocity dispersion, $1.6<\rm{log}_{10}(\sigma)<2.5$, and importantly both show the divergence between environmental classifications increasing towards low $\sigma$, an important result in our analysis.

\subsection{Physical implications}

In agreement with previous studies, we reproduce the well known linear relationship between $[\alpha/\text{Fe}]$ and $\sigma$, which is often taken to mean that star formation in massive galaxies was quenched relatively faster than in less massive galaxies.
However, we have also found substantial evidence to suggest that the $[\alpha/\text{Fe}]$ ratio has a strong dependence on the local environmental density.
Consistent with the observations of \cite{Hirschmann+14}, where the quiescent fraction increases with local environmental density, Fig.~\ref{fig:ETG_unconstrained_combined} shows that on average ETGs in HDEs are  $\alpha$-enhanced over their counterparts in LDEs.
Controlling for the correlation with stellar velocity dispersion, this appears to manifest primarily at low $\sigma$, and we stress that this simple division does not appear to extend to the high $\sigma$ regime.
From our own analysis, we infer that the integrated star-formation timescales cannot differ substantially between high-$\sigma$ galaxies across varied environments and morphologies, limiting the formation pathways.

This interpretation appears to support the work of \citet[][hereafter:\ \citetalias{Peng+10}]{Peng+10}, who found strong evidence for the environment and galaxy mass being independently associated with two distinct quenching processes.
Their work divided galaxies into ``red passive'' and ``blue star forming'', using the fraction of red galaxies as an indicator of the quenched fraction at a given environmental density and stellar mass.
Throughout this paper, we have interpreted galaxies with the highest values of $[\alpha/\text{Fe}]$ as having a rapidly quenched burst of star formation, and those with solar to sub-solar $\alpha$-enhancement as having ongoing star formation.
Hence, although $[\alpha/\text{Fe}]$ is not directly linked to the red fraction $f_{\rm{red}}$ used by \citetalias{Peng+10}, we consider it a useful comparison.

\begin{figure}
    \centering
    \includegraphics[width=\columnwidth]{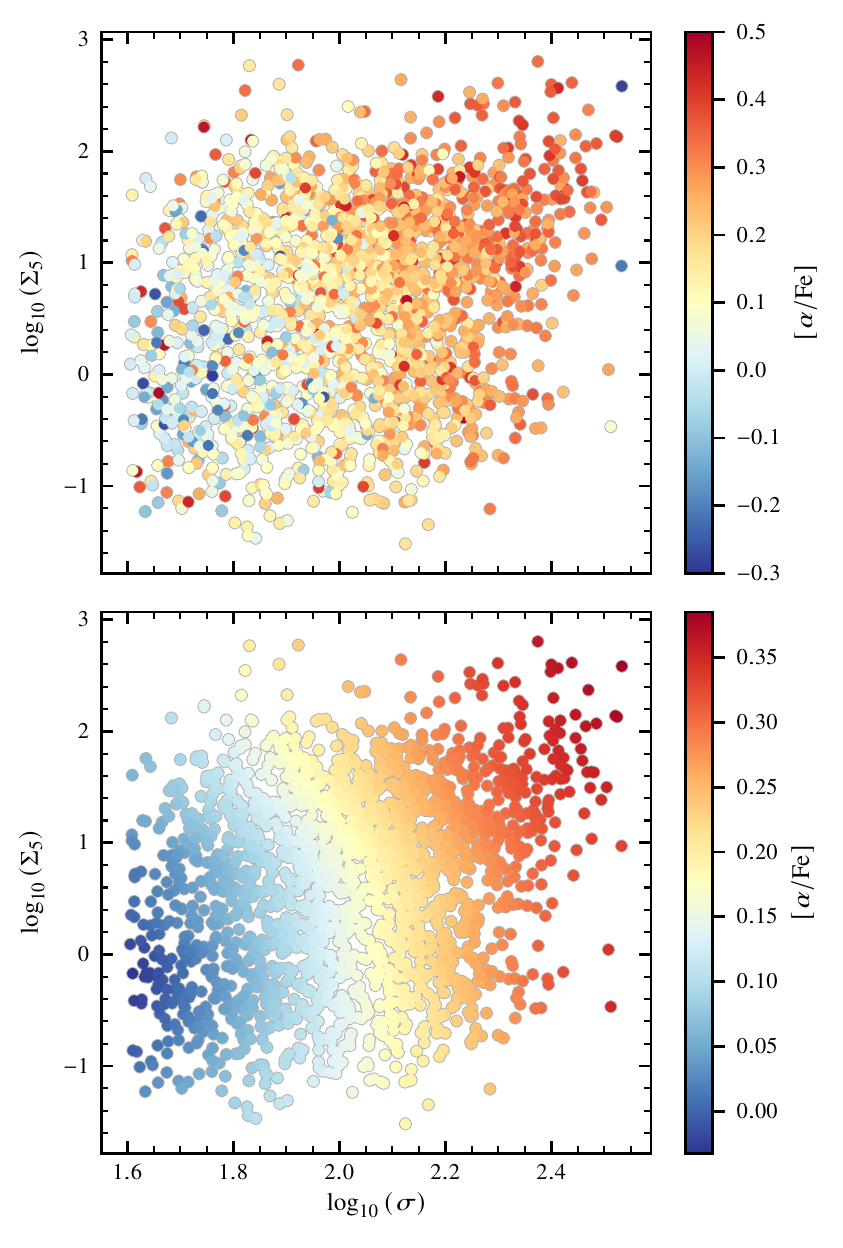}
    \caption[]{
    The distribution of galaxies in $\sigma$-$\Sigma_5$ space.
    In (a), galaxies are coloured by their associated measurements of $[\alpha/\text{Fe}]$.
    In (b), we show the same plot, but we use a locally smoothed value of $[\alpha/\text{Fe}]$, determined using the locally-weighted regression technique (LOESS) of \cite{Cappellari+13b}.
    This clearly show an underlying symmetry to the plane, with high-$\sigma$ galaxies in HDEs having the highest average values of $[\alpha/\text{Fe}]$.
    }
    \label{fig:peng_comparison}
\end{figure}

We show in Fig.~\ref{fig:peng_comparison} the distribution of $[\alpha/\text{Fe}]$ in the $\sigma$-$\Sigma_5$ plane.
Although the raw data shows a considerable degree of scatter, the locally smoothed values display a similar symmetry to the red fraction plotted in Fig.~6 of \citetalias{Peng+10}, with a line of constant $[\alpha/\text{Fe}]$ stretching from medium-$\sigma$ galaxies in LDEs, to low-$\sigma$ galaxies in HDEs.
Assuming that $[\alpha/\text{Fe}]$ is directly related to the quenching timescale in some manner, this simple symmetry appears to support the independence of both the mass and environmental quenching mechanisms.

We also consider the implications of Fig.~15 in \citetalias{Peng+10}, which suggests that the dominant quenching mechanism at low redshift depends on the stellar mass of the galaxy.
In particular, environmental quenching is predicted to be dominant for low-mass galaxies. 
This seems to be consistent with our results, wherein the $\alpha$-enhancement of low-$\sigma$ galaxies appears to be strongly influenced by the local environmental surface density, whereas only a faint imprint exists for galaxies at high-$\sigma$.
Further support for the mass dependence of the dominant quenching mechanism comes from considering the parabolic relationship in Fig.~\ref{fig:parabola}.
At $\rm{log}_{10}(\sigma)=1.8$, there is an enhancement of $0.069$\,dex in $[\alpha/\text{Fe}]$, which falls to just over half that at $\rm{log}_{10}(\sigma)=2.4$.
Thus, all elliptical galaxies show the imprint of environmental quenching, although it is still more pronounced at low velocity dispersion.
We suggest that our results also align with the na\"ive expectation, that the shallower potentials in low-$\sigma$ galaxies may make them more susceptible to environmental disruption.

Separately, we consider the flattening observed at high velocity dispersion.
The well-known mass-metallicity relation \citep{Lequeux+79} is also found to flatten towards higher stellar masses ($M_*>10^{10.5}\rm{M_{\odot}}$), and we suggest that this may have some correlation with the effect that we have observed.
This effect appears to be independent of environment, although strongly related to morphology.
Further research in this area is clearly needed, to confirm that the observed flattening is a physical effect, and to discern the extent to which it depends on morphology rather than velocity dispersion or stellar mass.

We estimate the effect of the low-$\sigma$ $\alpha$-enhancement using the empirical relationship from \cite{de_la_Rosa+11}, which relates $[\alpha/\text{Fe}]$ to the half-mass time $\rm{T}_{M/2}$ (the time interval over which half of the total stellar mass is formed).
For elliptical galaxies, the measured $\alpha$-enhancement of 0.057\,dex would imply $\rm{T}_{M/2}=(1.88\pm0.17)$\,Gyr for galaxies in HDEs, compared to $\rm{T}_{M/2}=(2.75\pm0.14)$\,Gyr for those in LDEs.  
Similarly, we also use \cite{de_la_Rosa+11} to estimate the stellar mass fraction contributed by stars older than 10\,Gyr, denoted as OLD.
For low-$\sigma$ ellipticals, we find OLD$_{\text{LDE}}=(56.8\pm1.3)\%$, and OLD$_{\text{HDE}}=(64.7\pm1.5)\%$.
There appears to be a strong residual dependence on morphology here as well, with the environmental offsets greatest in ellipticals, and statistically insignificant in Sc/Irr galaxies.
We hypothesise that this may be due in part to the effects of ongoing star formation, since a clear picture of the star-formation timescale would be obscured by any more recent starbursts.

\section{Conclusions} \label{sec:conclusions}

We have made a new set of measurements of $[\alpha/\text{Fe}]$ for all galaxies in the SAMI Galaxy Survey.
These are an improvement over previous studies, primarily by utilising the variance-weighted method of \cite{Cenarro+01} to measure the underlying Lick/IDS indices.
Based on our subsequent analysis, we make the following observations:
\begin{enumerate}
    \item Separating galaxies by their $k$-corrected $(g-i)$ colour, we find red sequence galaxies are consistently $\alpha$-enhanced over those in the blue cloud, at fixed velocity dispersion $\sigma$. The relationship for red sequence galaxies can be described by the equation  $$[\alpha/\text{Fe}]=(0.378\pm0.009)\rm{log}_{10}\left(\sigma/100\right)+(0.155\pm0.003).$$
    \item Taking the residuals $\Delta[\alpha/\text{Fe}]$ from a linear relationship with $\sigma$, and investigating the variation with the local surface density, $\Sigma_5$, we find a weak correlation for red sequence galaxies, $$\Delta[\alpha/\text{Fe}]=(0.021\pm0.003)\rm{log}_{10}(\Sigma_5)-(0.009\pm0.003).$$
    In contrast, no such correlation exists for blue cloud galaxies.
    \item At low velocity dispersion, $\sigma<100$\;km\,s$^{-1}$, and controlling for morphology, we find that galaxies in high-density environments ($\rm{log}_{10}(\Sigma_5)>1.15$) are $\alpha$-enhanced by up to $\sim$0.06\,dex compared to their counterparts in low-density environments ($\rm{log}_{10}(\Sigma_5)\leq1.15$).
    \item This enhancement includes a strong morphological dependence, with the offset increasing from $0.019\pm0.010$\,dex in spirals to $0.057\pm0.014$\,dex in ellipticals. 
    Using the models of \cite{de_la_Rosa+11}, we estimate the inferred difference in half-mass formation time for ellipticals as $\Delta \rm{T}_{\rm{M/2}}=0.87\pm0.22$\,Gyr.
    \item Controlling for the environment, we find some indication of an additional morphological component. Considering galaxies in high-density environments at $\rm{log}_{10}(\Sigma_5)=1.6$, we expect ellipticals to be $\alpha$-enhanced by 0.09\,dex over S0s, and by 0.13\,dex over Sa/b galaxies. The evidence for this is weaker than the environmental effect, and would benefit greatly from further study.
    \item There is evidence to suggest a flattening of the $[\alpha/\text{Fe}]$-$\sigma$ relation at high velocity dispersions in elliptical galaxies.
    This flattening exists independently of the environment.
    However, we find no support for such a relationship for later morphological types.
\end{enumerate}

\section*{Acknowledgements}

The  SAMI Galaxy Survey is based on observations made at the Anglo-Australian Telescope, and was developed jointly by the University of Sydney and the Australian Astronomical Observatory.
The SAMI input catalogue is based on data taken from the Sloan Digital Sky Survey, the GAMA Survey, and the VST ATLAS Survey.
The SAMI Galaxy Survey is supported by the Australian Research Council Centre of Excellence for All Sky Astrophysics in 3 Dimensions (ASTRO 3D), through project number CE170100013, the Australian Research Council Centre of Excellence for All-sky Astrophysics (CAASTRO), through project number CE110001020, and other participating institutions. 
The SAMI Galaxy Survey website is \href{http://sami-survey.org/}{http://sami-survey.org/}.

PJW and RLD acknowledge travel and computer grants from Christ Church, Oxford and support from the Oxford Hintze Centre for Astrophysical Surveys which is funded by the Hintze Family Charitable Foundation. 
RLD is also supported by the Science \& Technology Facilities Council grant numbers ST/H002456/1, ST/K00106X/1 and ST/J002216/1.
SB acknowledges funding support from the Australian Research Council through a Future Fellowship (FT140101166).
FDE acknowledges funding through the H2020 ERC Consolidator Grant 683184.
BG is the recipient of an Australian Research Council Future Fellowship (FT140101202). 
JvdS acknowledges support of an Australian Research Council Discovery Early Career Research Award (project number DE200100461) funded by the Australian Government. 
NS acknowledges support of an Australian Research Council Discovery Early Career Research Award (project number DE190100375) funded by the Australian Government and a University of Sydney Postdoctoral Research Fellowship. 
JBH is supported by an ARC Laureate Fellowship FL140100278. 
The SAMI instrument was funded by Bland-Hawthorn's former Federation Fellowship FF0776384, an ARC LIEF grant LE130100198 (PI Bland-Hawthorn) and funding from the Anglo-Australian Observatory. 
JJB acknowledges support of an Australian Research Council Future Fellowship (FT180100231). 
MSO acknowledges funding support from the Australian Research Council through a Future Fellowship (FT140100255). 

\section*{Data Availability}

All data used in this work are publicly available, through the  \href{https://docs.datacentral.org.au/sami}{SAMI Data Release 3} \citep{Croom+21}.



\bibliographystyle{mnras}
\bibliography{full_list.bib} 




\appendix

\section{Index Measurements}
\label{app:index_measurements}

We present a more detailed explanation of the process of measuring the line indices.
We treat the index as an equivalent width, defined over three wavelength bandpasses. 
We take $\lambda_{b_1}$ and $\lambda_{b_2}$ as the wavelength limits of the blue bandpass and similarly $\lambda_{r_1}$ and $\lambda_{r_2}$ as the limits of the red bandpass. 
These are used to derive the local pseudo-continuum, $C(\lambda)$, through an error weighted least-squares fit.
We begin by writing the continuum level as 
\begin{equation} \label{eq:cenarro_cont_level_1}
C(\lambda) = \alpha_1 + \alpha_2 \lambda,
\end{equation}
where we use the following definitions from \cite{Cenarro+01}
\begin{equation} \label{eq:cenarro_cont_level_2}
    \alpha_1 = \frac{1}{\Delta} \left( \Sigma_3 \Sigma_4 - \Sigma_2 \Sigma_5 \right),
\end{equation}
\begin{equation} \label{eq:cenarro_cont_level_3}
    \alpha_2 = \frac{1}{\Delta} \left( \Sigma_1 \Sigma_5 - \Sigma_2 \Sigma_4 \right),
\end{equation}
\begin{equation} \label{eq:cenarro_cont_level_4}
    \Delta = \Sigma_1 \Sigma_3 - \Sigma_2 \Sigma_2,
\end{equation}
with the parameters
\begin{equation} \label{eq:cenarro_cont_sigma_1}
    \Sigma_1 \equiv \sum_{n=1}^{N_c} \sum_{h=1}^{M(n)} \frac{1}{\sigma^2 \left[ S ( \lambda_{n,h} ) \right] },
\end{equation}
\begin{equation} \label{eq:cenarro_cont_sigma_2}
    \Sigma_2 \equiv \sum_{n=1}^{N_c} \sum_{h=1}^{M(n)} \frac{\lambda_{n,h}}{\sigma^2 \left[ S ( \lambda_{n,h} ) \right] },
\end{equation}
\begin{equation} \label{eq:cenarro_cont_sigma_3}
    \Sigma_3 \equiv \sum_{n=1}^{N_c} \sum_{h=1}^{M(n)} \frac{\lambda_{n,h}^2}{\sigma^2 \left[ S ( \lambda_{n,h} ) \right] },
\end{equation}
\begin{equation} \label{eq:cenarro_cont_sigma_4}
    \Sigma_4 \equiv \sum_{n=1}^{N_c} \sum_{h=1}^{M(n)} \frac{S ( \lambda_{n,h} )}{\sigma^2 \left[ S ( \lambda_{n,h} ) \right] },
\end{equation}
\begin{equation} \label{eq:cenarro_cont_sigma_5}
    \Sigma_5 \equiv \sum_{n=1}^{N_c} \sum_{h=1}^{M(n)} \frac{\lambda_{n,h} S ( \lambda_{n,h} ) }{\sigma^2 \left[ S ( \lambda_{n,h} ) \right] },
\end{equation}
defined following Cenarro's notation, where we sum over the $M(n)$ pixels in $N(c)$ continuum bandpasses. 
In our case, we only use one red and one blue bandpass to derive the continuum level, following the definitions of the Lick indices by \cite{Worthey+97}. 
We use $\sigma^2 \left[ S ( \lambda_{n,h} ) \right] $ to denote the variance of the observed spectrum at the central wavelength of the $h^{\text{th}}$ pixel in the $n^{\text{th}}$ continuum bandpass. 
In addition, we note that at the borders of the continuum bandpasses the wavelengths will not correspond exactly to the edges of the outer pixels, and as such, contributions from fractional pixels must be included in the summations. 

However, we differ from Cenarro for the central index bandpass, where we eschew this pixel-based summation over the central index bandpass in favour of numerical integration.
We use the definition of a classical atomic index,
\begin{equation} \label{eq:classic_index_atomic}
I_a(\text{\AA})\equiv\int_{\lambda_{c_1}}^{\lambda_{c_2}}\left[1-\frac{S(\lambda)}{C(\lambda)}\right]\text{d}\lambda,
\end{equation}
where $S(\lambda)$ is the flux spectrum, and we integrate between $\lambda_{c_1}$ and $\lambda_{c_2}$, the limits of the central index feature bandpass.
For the integral, we interpolate over the flux spectrum $S(\lambda)$ using a piecewise linear spline, and numerically integrate between the precise wavelength limits of the central bandpass.
Similarly, for molecular indices measured in magnitudes, we define
\begin{equation} \label{eq:classic_index_molecular}
    I_m(\text{mag}) \equiv -2.5 \log_{10} \left( 1 - \frac{I_a}{\lambda_{c_2} - {\lambda_{c_1}}} \right).
\end{equation}

\section{Sample selection bias} 
\label{app:sample_selection_bias}

\subsection{Redshift offsets}
\label{app:redshift_offsets}

\begin{figure}
    \centering
    \includegraphics[width=\columnwidth]{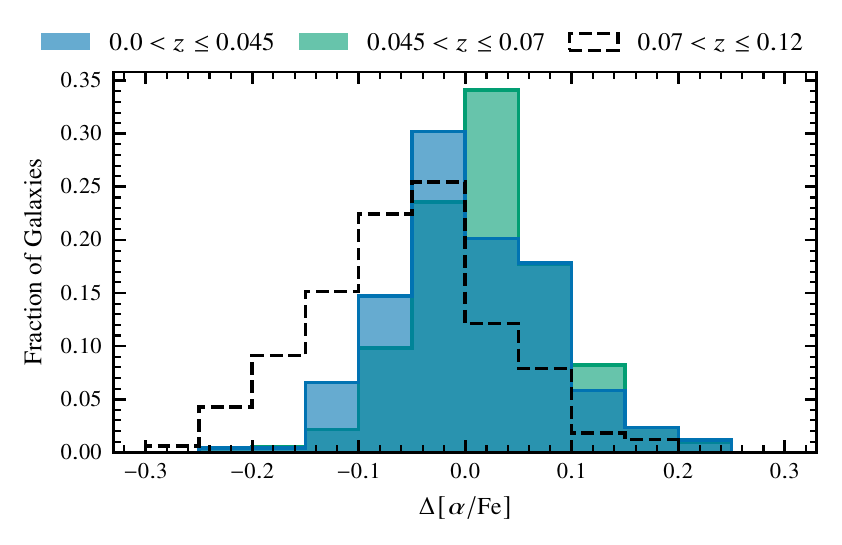}
    \caption[]{The residuals of our ETG sample from a best fit linear trendline, separated by redshift.
    A shift to lower values of $[\alpha/\text{Fe}]$ is clearly visible for the highest redshift bin.
    The offset of the median of the higher redshift distribution is $\sim0.065$\,dex, a similar order of magnitude to the effects we investigate in this study.
    }
    \label{fig:ETG_redshift_residuals_subplots_2}
\end{figure}

\begin{figure}
    \centering
    \includegraphics[width=\columnwidth]{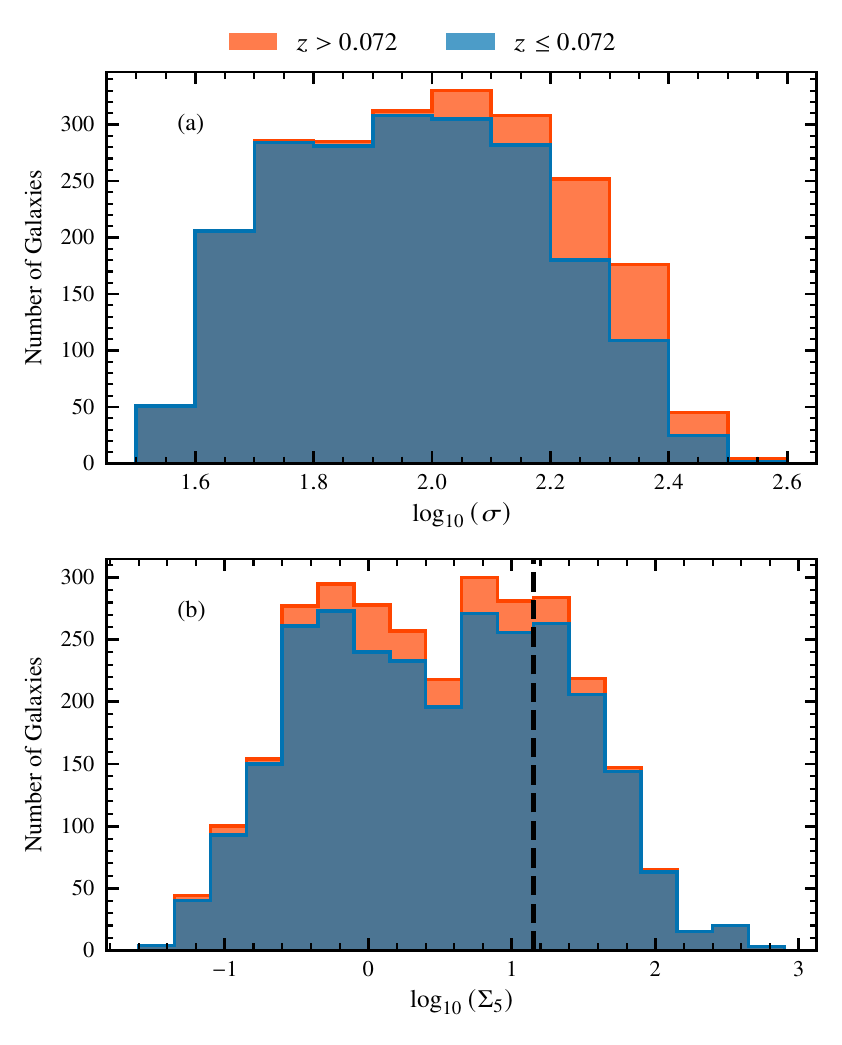}
    \caption[]{The distribution of galaxies in the SGS, separated by redshift. 
    Galaxies with $z>0.072$ are clearly biased towards higher velocity dispersions, shown in (a), and are predominantly found at lower local surface densities, displayed in (b). 
    In (b), the dashed line represents the value of $\rm{log}_{10}\left(\Sigma_5\right)$ chosen to separate galaxies into low- and high-density environments.
    }
    \label{fig:redshift_hist_sig_surf_dens}
\end{figure}

As mentioned in Section \ref{subsec:spectral_fitting}, all galaxies in the SGS with $z>0.072$ were removed from the sample.
This was due to contamination of the spectrum from the \ion{O}{i} atmospheric skyline, at 5577\,\AA.
At a redshift $z=0.072$, the centroid of this line is coincident with the edge of the central bandpass for Mg$_2$.
The SSP models we have used are very sensitive to changes in the Magnesium indices, particularly for determining $[\alpha/\text{Fe}]$.
We show an example of the effect of this in Fig.~\ref{fig:ETG_redshift_residuals_subplots_2}, where for all ETGs, we fit $[\alpha/\text{Fe}]$ to a common relationship against $\sigma$, and plot the residuals as a function of redshift.
The clear offset for the higher redshift galaxies is concerning, but we also note that these are exclusively higher-mass field and group galaxies drawn from the GAMA survey \citep{Driver+11}, with the relative distributions of $\sigma$ and $\Sigma_5$ shown in Figure \ref{fig:redshift_hist_sig_surf_dens}.
The higher-redshift galaxies would therefore unfairly bias the sample at lower environmental densities to lower values of $[\alpha/\text{Fe}]$.
Since a primary component of our study is the comparison of galaxies across environments, we elected to remove these galaxies from consideration. 
It is possible that this effect could be mitigated in future by the use of alternative methods for determining the stellar populations, such as different SSP models, or utilising full-spectrum fitting.
Although this cut will have a measurable influence on our analysis, we consider it necessary to avoid the possibility of a more pronounced bias, as shown in Figure \ref{fig:ETG_redshift_residuals_subplots_2}.

\subsection{Model limitations}
\label{app:model_limitations}

\begin{figure}
    \centering
    \includegraphics[width=\columnwidth]{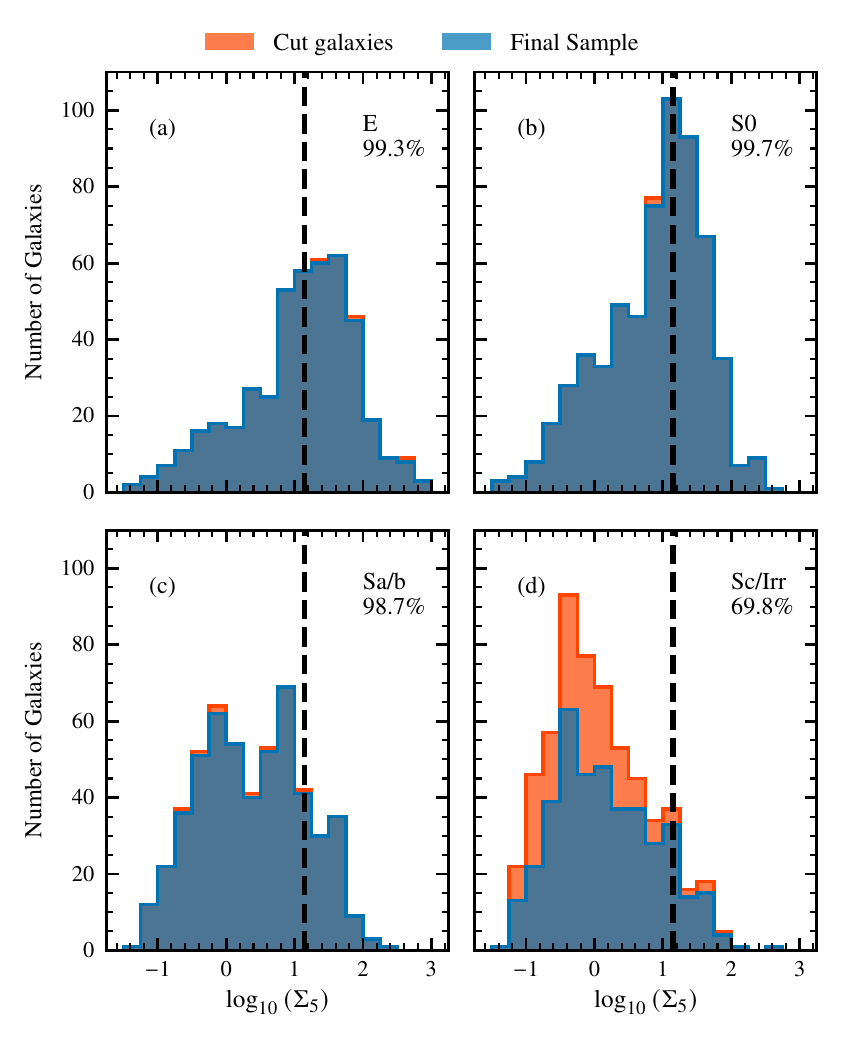}
    \caption[]{
    We show the effects of cutting galaxies due to parameter space limitations on $[\alpha/\text{Fe}]$. The completeness of the sample is relative to before and after this particular quality cut, with the overall numbers of galaxies available in Table \ref{tab:total_number_of_galaxies}.
    It is clear that this cut predominantly affects Sc/Irr galaxies, with a bias to low local surface densities.}
    \label{fig:alpha_cut_galaxies}
\end{figure}

In comparison to the other quality cuts, the decision to remove galaxies based on their measured values of $[\alpha/\text{Fe}]$ has the potential to bias the resulting sample, if the set of galaxies removed correlates with other properties.
For our study, we removed galaxies where $[\alpha/\text{Fe}]$ converged to the limits of the parameter space, or where the uncertainty $\delta[\alpha/\text{Fe}]$ spanned the entire space.
We compare the original and resultant distributions in Figure \ref{fig:alpha_cut_galaxies}, as a function of morphology and local surface density.

For E, S0, and Sa/b galaxies, the sample completeness considering only this cut is in excess of 98\%, and therefore we do not anticipate this measure introducing any significant bias.
In contrast, for galaxies classified as Sc/Irr, we have removed $\sim$30\% of the sample, with the effect concentrated around $\rm{log}_{10}(\Sigma_5)=0$.

\section{Measurement comparison}
\label{app:measurement_comparison}

\begin{figure}
    \centering
    \includegraphics[width=\columnwidth]{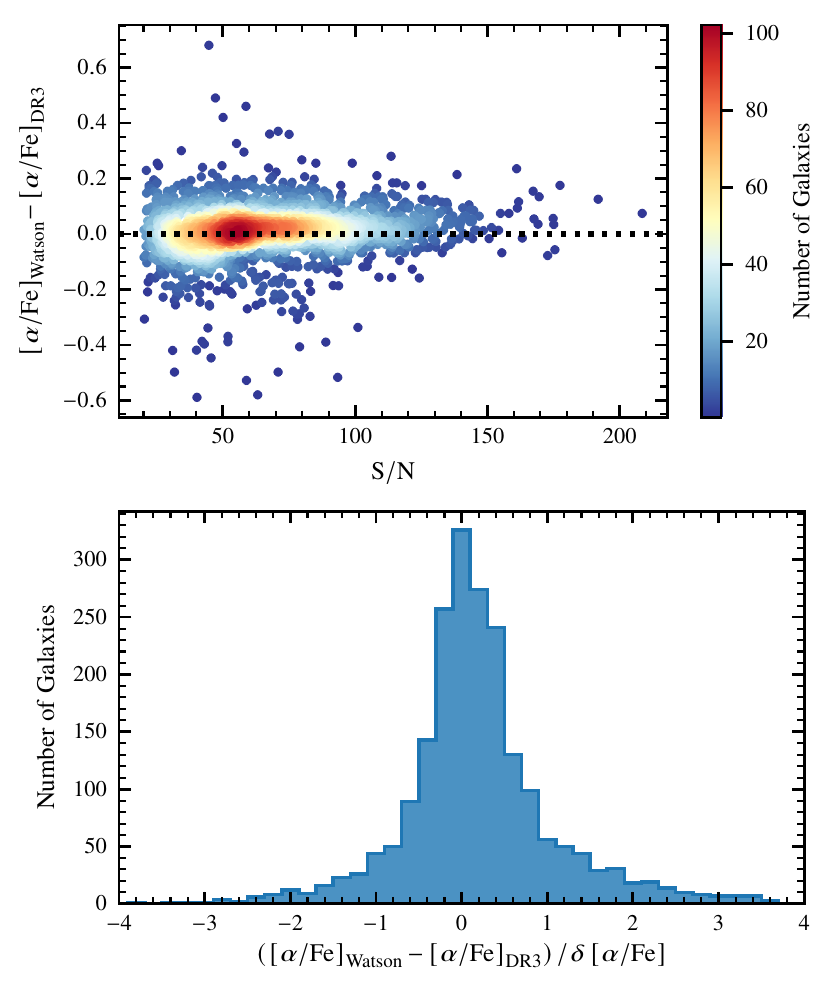}
    \caption[]{
    The offset between the measurements of $[\alpha/\text{Fe}]$ in this work, denoted by  $[\alpha/\text{Fe}]_{\rm{Watson}}$, and those included in SAMI DR3, $[\alpha/\text{Fe}]_{\rm{DR3}}$.
    We display in (a) the offset as a function of the spectral signal-to-noise ratio.
    Galaxies are shaded according to the local number density, or the number of points within each grid space (10\,dex in S/N, and 0.05\,dex in $\Delta[\alpha/\text{Fe}]$), to account for the high degree of overplotting.
    The dashed line is included purely for reference purposes, and delineates a zero offset between the two samples.
    In (b), the offset is normalised by the associated error, and we show the resulting distribution.
    Although the distribution shows wider wings than a standard gaussian, it is also more sharply peaked.
    }
    \label{fig:DR3_comparison}
\end{figure}

We show a comparison between our measurements and those released as part of SAMI DR3 in Figure \ref{fig:DR3_comparison}.
We compare only those 2093 galaxies included in our final sample, to reduce the effect of outliers which would not be considered in any analysis.
A visual inspection indicates a good agreement between the two versions, which is reinforced by the underlying data.
The scatter in the offset increases substantially towards lower S/N, although we do not observe any significant shift in the distribution, which remains centred around approximately zero offset.
Considering the normalised offset, $([\alpha/\text{Fe}]_{\rm{Watson}}-[\alpha/\text{Fe}]_{\rm{DR3}})/\delta[\alpha/\text{Fe}]$ (denoted as $x$ for simplicity), we find our measurements are in fact shifted to slightly higher values of  $[\alpha/\text{Fe}]$, with the median of the distribution being 0.11$x$.
We calculate the standard deviation of the distribution as 1.1$x$, although we also highlight that 79\% of galaxies fall within 1$x$, more than would be expected if the distribution followed a standard gaussian.
We therefore consider our results in excellent agreement with the preceding measurements at a population level.





\bsp	
\label{lastpage}
\end{document}